\documentclass[journal=jctcce,manuscript=article]{achemso}

\usepackage[version=3]{mhchem} 
\usepackage[dvipsnames]{xcolor}
\usepackage{bm}
\usepackage{threeparttable}
\usepackage{amsmath,amsfonts,amssymb}
\usepackage{textcomp,gensymb}
\usepackage{booktabs}
\usepackage{graphicx,caption}
\usepackage{microtype}
\usepackage{soul}
\usepackage[symbol]{footmisc}

\author{Xinze Li}
\affiliation{Department of Chemistry, Shanghai Key Laboratory of Electrochemical and Thermochemical Conversion for Resources Recycling, Fudan University, Shanghai, 200438, China}
\author{Ruitao Ma}
\affiliation{Department of Chemistry, Shanghai Key Laboratory of Electrochemical and Thermochemical Conversion for Resources Recycling, Fudan University, Shanghai, 200438, China}
\author{Chen Qu}
\affiliation{Independent Researcher, Toronto, Ontario M9B0E3, Canada}
\author{Dong H. Zhang}
\affiliation{State Key Laboratory of Molecular Reaction Dynamics, Dalian Institute \\of Chemical Physics, Chinese Academy of Sciences, Dalian, 116023, China}
\author{Qi Yu}
\email{qi_yu@fudan.edu.cn}
\affiliation{Department of Chemistry, Shanghai Key Laboratory of Electrochemical and Thermochemical Conversion for Resources Recycling, Fudan University, Shanghai, 200438, China}

\title{Monomeric machine learning potential for general covalent molecules: linear alkanes as an example}

\begin{document}

\newpage

\begin{abstract}

Machine-learning potentials (MLPs) have become important tools for modern molecular simulations. However, developing models that simultaneously achieve high accuracy and high computational efficiency remains a significant challenge. In this work, we extend the recently proposed MB-PIPNet framework to general covalently bonded molecular systems by combining monomer-based energy decomposition, permutationally invariant polynomial (PIP) descriptors, and neural networks within a fragmentation-based strategy. Within this framework, the total potential energy is represented as a sum of effective monomeric contributions, where PIPs provide compact and chemically motivated descriptions of both monomer internal structures and their local chemical environments. As a proof-of-concept application, we apply the MB-PIPNet framework to linear alkanes, using \ce{C14H30} as a representative system, and benchmark its performance against established atomistic machine-learning models. The resulting MB-PIPNet potential accurately reproduces reference \emph{ab initio} electronic energies and reliably captures key molecular properties, including torsional potential energy profiles, harmonic vibrational frequencies, and vibrational power spectra obtained from molecular dynamics simulations. Importantly, MB-PIPNet demonstrates a substantial advantage in computational efficiency over other MLP models for combined energy and force evaluations. These results establish MB-PIPNet as a scalable and efficient framework for constructing MLPs, providing an additional route for large-scale quantum and classical simulations of complex molecular systems.
\end{abstract}   


\newpage

\section{Introduction}
Over the past two decades, machine learning potentials (MLPs) have emerged as a powerful and promising tool for enabling efficient yet accurate simulations of complex chemical, biological, and materials processes.\cite{gkeka2020machine,deringer2019machine,NN-2014,manzhos2020neural,meuwly2021machine,Braams2009,ARPC2018,PIP-NN-1,Guo16,FINN1,fu2023accurate,BPNN,behler2021four,chmiela2018towards,bartok2010gaussian,GP-2017-1,SchNet,PhysNet,DeepMD,EANN,NequIP,MACE,Allegro,jiang2025review}  A broad class of modern ML approaches adopts the Behler–Parrinello–type atomistic representation,\cite{BPNN} in which the total potential energy of a system is expressed as a sum of atomic local energy contributions. Some representative invariant and equivariant neural network potentials developed within this framework include BPNN,\cite{BPNN} SchNet,\cite{SchNet} PhysNet,\cite{PhysNet} DeePMD,\cite{DeepMD} EANN,\cite{EANN} NequIP,\cite{NequIP} MACE,\cite{MACE} and Allegro.\cite{Allegro} The widespread adoption of atomistic energy decomposition can largely be attributed to its general applicability across reactive and non-reactive molecular systems as well as extended materials. In addition, this strategy leads to a computational cost that scales linearly with the number of atoms, making it particularly attractive for high-dimensional systems. Despite these advantages, the assumption of atomic local energies lacks direct chemical interpretability, as molecular structure and dynamics are governed by the total molecular energy rather than by individual atomic contributions.\cite{heidar2017information,uhligEmergenceAccurateAtomic2025,FFLUX,QCTo1} Furthermore, although linear scaling is computationally favorable, it is not necessarily optimal, and opportunities remain for further reductions in computational complexity.

As another class of MLP approaches, permutationally invariant polynomials (PIPs) have also been extensively developed and applied.\cite{Braams2009,Xie10,ARPC2018} PIPs construct polynomial descriptors of molecular systems using internuclear distances or, more commonly, Morse-type variables, thereby ensuring exact invariance with respect to translation, rotation, and permutation. This has enabled the development of highly accurate potential energy surfaces (PESs) for molecules,\cite{Bowman2011,ARPC2018,czako2021,BullVulpe2023} molecular clusters,\cite{purified15c,Chen15,yujcp} water systems,\cite{WHBB,mbpoltests,q_AQUA,Mbpol23,qAQUApol} and, more recently, condensed-phase materials.\cite{aPIP2020} PIPs and their minimal subsets, known as fundamental invariants (FIs),\cite{FINN1,Fu18,fu2023accurate} have also been successfully employed as input descriptors for neural-network-based models\cite{Guo16,Fu18,fu2023accurate} and Gaussian process regression\cite{PIP-GP} potentials. For a comprehensive overview of the theoretical foundations and recent developments of PIP-based approaches, interested readers are referred to our recent review article.\cite{bowman2025perspective}

When combined with many-body expansion (MBE), PIP-based approaches have been successfully employed to construct highly accurate and transferable PESs for both non-covalent and covalent molecular systems.\cite{WHBB,purified15a,mbpoltests,MBfit,MBX1.2,q_AQUA,zhangFIMB2022,Mbpol23,BullVulpe2023} Within this framework, PIP-based PESs are fitted to high-level \emph{ab initio} reference data for individual $n$-body interaction terms, including the one-body ($1$-b), two-body ($2$-b), three-body ($3$-b), and higher-order contributions, where ``b" denotes the molecular monomer. Notable examples include the ``gold-standard" CCSD(T)-quality water potentials q-AQUA,\cite{q_AQUA} q-AQUA-pol,\cite{qAQUApol} and MB-pol(23).\cite{Mbpol23} Despite their remarkable success in quantitatively describing both gas-phase clusters and condensed-phase systems, PIP-based methods combined with the MBE face intrinsic scalability challenges. In particular, the construction of PIP bases becomes increasingly expensive with molecular size, while the number of $n$-body interaction terms grows exponentially for increasing order of expansion. 

Recently, we introduced a new machine-learning framework, MB-PIPNet, which integrates the strengths of atomistic MLPs, PIPs, and the many-body expansion, while mitigating several of their respective limitations.\cite{yu2025extending} In MB-PIPNet, the total potential energy of a molecular system is decomposed into the sum of monomeric energies, rather than atomic local energies. For each molecular monomer, the perturbed monomeric energy is represented by a feed-forward neural network that takes PIP-based structural descriptors as input. These descriptors are constructed to capture both the intramolecular geometry of the monomer and its effective two-body interactions with the surrounding environment. The MB-PIPNet framework has been systematically validated for systems ranging from gas-phase molecular clusters to condensed-phase liquid water and \ce{CO2}. The resulting PESs were shown to achieve high accuracy while maintaining favorable computational efficiency, owing to the use of only one-body and two-body PIP descriptors and a computational cost that scales linearly with the number of monomers. To date, however, these applications have been restricted to noncovalent molecular systems, for which the definition of monomers is unambiguous. The extension of MB-PIPNet to covalently bonded systems therefore remains an open and unexplored problem.

In this work, we present a further extension of the MB-PIPNet framework to covalent systems through the usage of a fragmentation-based scheme.\cite{qu2019fragmented,conte20,BullVulpe2023,qu2024hydrocarbon,qu2024dynamics} This strategy builds upon the concepts of the many-body expansion and related molecular fragmentation approaches,\cite{zhang2003molecular,Collins2005approximate,collins2007molecular,Gadre2014molecular} and systematically incorporates them into the MB-PIPNet formalism at the two-body level. As a proof-of-concept, we consider long linear alkane molecules, specifically \ce{C14H30}, and demonstrate that the resulting MB-PIPNet PES accurately captures the essential interactions in such covalently bonded systems, while retaining significantly lower computational cost compared to other MLPs.
 
The remainder of this paper is organized as follows. First, we briefly review the general MB-PIPNet framework. We then describe in detail the fragmentation scheme and its integration into MB-PIPNet for covalent molecules, using linear alkane \ce{C14H30} as an example. Computational details are provided for the construction of the density functional theory (DFT) level training and test datasets, the MB-PIPNet model architecture and training protocol, and the molecular dynamics simulations. The Results section presents benchmark assessments of the resulting MB-PIPNet PES for \ce{C14H30}, including predictions of energies, typical potential energy curves, harmonic vibrational frequencies, and power spectra, as well as an analysis of the computational efficiency in comparison with other MLPs. Finally, we summarize the current work and discuss future broader applications of the MB-PIPNet framework.

\section{Theory and Computational Details}

\subsection{The MB-PIPNet Approach}

\begin{figure}[H]
\begin{center}
\includegraphics[width=\textwidth]{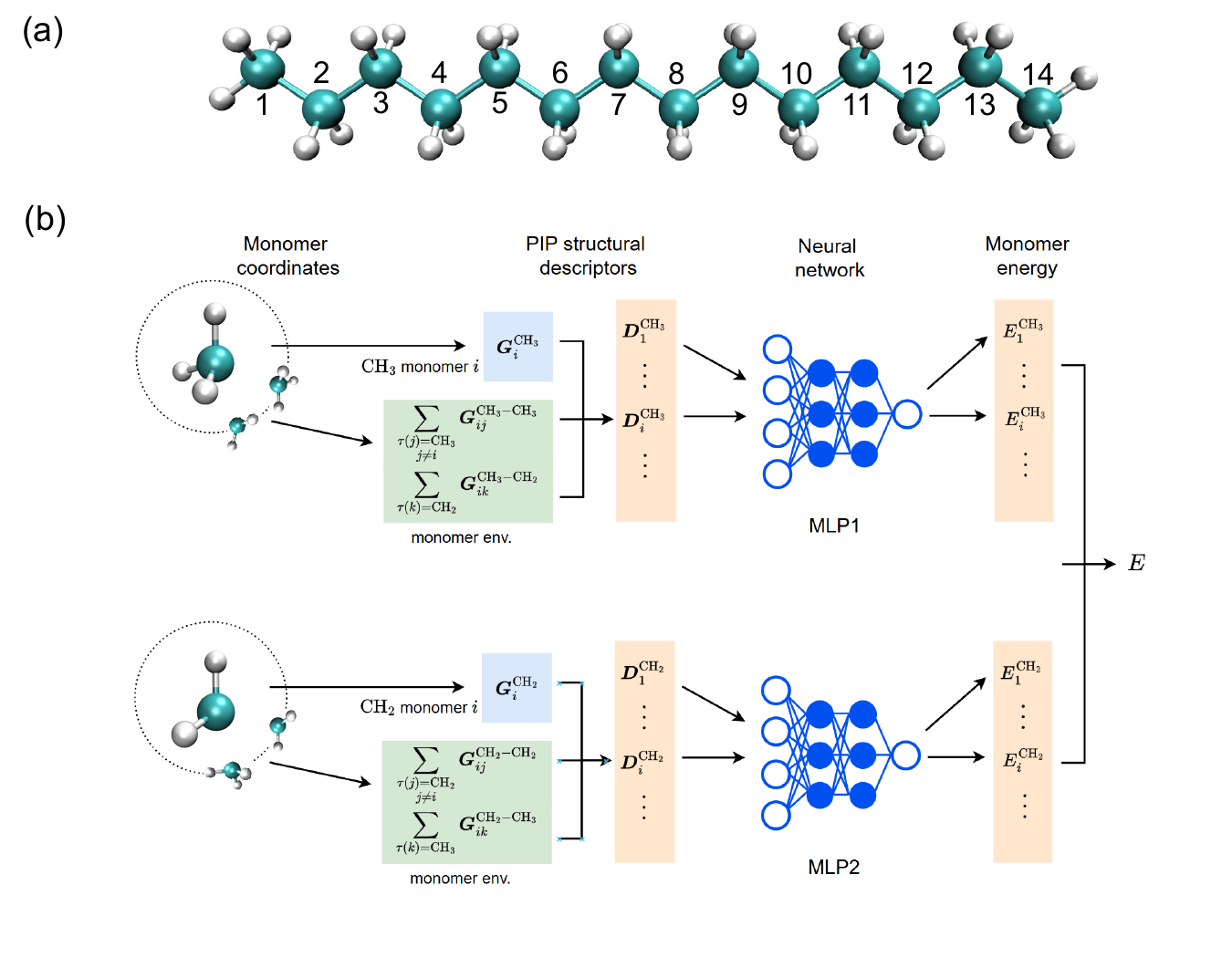}
\end{center}
\caption{a) Global minimum structure of the alkane \ce{C14H30}, with carbon atoms numbered. b) Schematic of the MB-PIPNet architecture with the example of linear alkane.}
\label{fig1}
\end{figure}

We first present the MB-PIPNet framework, detailing its fundamental energy decomposition scheme and the subsequent construction of the neural-network architecture. The application of this framework to covalent systems is illustrated in Fig. \ref{fig1}, using n-tetradecane \ce{C14H30} as a representative example.

As described in our previous work,\cite{yu2025extending} the total potential energy of a molecular system comprising $N$ atoms is divided into a sum of effective monomeric contributions from $N_\text{mon}$ monomers:

\begin{equation}
\begin{aligned}
E_{\text{total}} = \sum_{i=1}^{N_{\text{mon}}} E_i 
\end{aligned}
\end{equation}

The effective energy of monomer $i$ is calculated by
\begin{equation}
\begin{aligned}
E_i = \mathcal{E}_{\tau(i)}(\boldsymbol{D}_i)
\end{aligned}
\end{equation}
where $\tau(i)$ identifies the chemical species of monomer $i$, and the function $\mathcal{E}_{\tau(i)}$ represents the species-specific neural network that maps the local structural descriptor, $\boldsymbol{D}_i$, to the monomeric energy.

The local structural descriptor $\boldsymbol{D}_i$ incorporates both the self-structural features and environmental terms.

\begin{equation}
\begin{aligned}
\boldsymbol{D}_i = \boldsymbol{G}_{i}^{(1)} \oplus \sum_{j \neq i} \boldsymbol{G}_{ij}^{(2)}
\end{aligned}
\end{equation}

The self-structural descriptor, $\boldsymbol{G}_{i}^{(1)}$, characterizes the internal geometry of monomer $i$ and is composed of PIPs for the atomic set $\mathcal{A}_i$:

\begin{equation}
\begin{aligned}
\boldsymbol{G}_i^{(1)} = \hat{S}\left[ \prod_{a,b \in \mathcal{A}_i, a<b} p_{ab}^{l_{ab}} \right] 
\end{aligned}
\end{equation}
where $p_{ab}$ is Morse-like variable of the internuclear distance $r_{ab}$,
\begin{equation}
\begin{aligned}
p_{ab} = e^{-r_{ab}/\lambda}
\end{aligned}
\end{equation}
with a range hyperparameter $\lambda$. $l_{ab}$ is the order of the monomial, and $\hat{S}$ is the symmetrization operator that enforces permutational invariance for the identical atoms within the monomer.

The environmental descriptor, $\boldsymbol{G}_{ij}^{(2)}$, encodes the inter-monomeric interaction between monomer $i$ and its neighboring monomer $j$:

\begin{equation}
\begin{aligned}
\boldsymbol{G}^{(2)}_{ij} = \hat{S} \left[ \prod_{a,b \in \mathcal{A}_i \cup \mathcal{A}_j, a<b} {p}_{ab}^{l_{ab}} \right]
\end{aligned}
\end{equation}
where $\mathcal{A}_i \cup \mathcal{A}_j$ denotes the union of atoms in monomers $i$ and $j$. The hyperparameter $\lambda$ in $p_{ab}$ of $\boldsymbol{G}^{(2)}$ is usually distinct from that of $\boldsymbol{G}^{(1)}$, to capture two-body interactions more accurately.

As shown in Fig. \ref{fig1}a and \ref{fig1}b, in the \ce{C14H30} system, monomers are classified into two types: methyl (\ce{-CH3}) and methylene (\ce{-CH2-}). For the methyl unit, the structural descriptor can be constructed as

\begin{equation}
\begin{aligned}
\boldsymbol{D}_i^{\text{CH}_3} = \boldsymbol{G}_{i}^{\text{CH}_3} \oplus \sum_{\substack{\tau(j) = \text{CH}_3\\j \neq i}} \boldsymbol{G}_{ij}^{\text{CH}_3 \text{-CH}_3} \oplus \sum_{\tau(k) = \text{CH}_2} \boldsymbol{G}_{ik}^{\text{CH}_3 \text{-CH}_2}
\end{aligned}
\label{eq:dch3}
\end{equation}

For the methylene unit, the descriptor is

\begin{equation}
\begin{aligned}
\boldsymbol{D}_i^{\text{CH}_2} = \boldsymbol{G}_{i}^{\text{CH}_2} \oplus \sum_{\substack{\tau(j) = \text{CH}_2\\j \neq i}} \boldsymbol{G}_{ij}^{\text{CH}_2 \text{-CH}_2} \oplus \sum_{\tau(k) = \text{CH}_3} \boldsymbol{G}_{ik}^{\text{CH}_3 \text{-CH}_2}
\end{aligned}
\label{eq:dch2}
\end{equation}

The scheme of fragmentation into monomeric units is not unique. Other fragmentation schemes in electronic structure calculations\cite{gordon2012fragmentation,fragzhang,fragcollins} and PIP-based MLP constructions\cite{QuBowman2019,BullVulpe2023} can also be applied with appropriate adjustment.

\subsection{The MB-PES approach}
One of the authors recently reported a MLP for linear alkanes based on the many-body PIPs approach.\cite{jctctwists} This approach is used to refit the potential with the same training data from this work, and the resulting PES, denoted MB-PES, is compared to the MB-PIPNet potential.

The theory of the many-body PIP approach has been presented in detail in ref. \citenum{jctctwists}, and is described briefly here. Consider the usual many-body expansion of the total energy to 4-body interactions:
\begin{equation}
\begin{aligned}
    V = & \sum_i V^{(1b)}_i + \sum_{i,j} V^{(2b)}_{ij} + 
    \sum_{i,j,k}V^{(3b)}_{ijk} + \sum_{i,j,k,l}V^{(4b)}_{ijkl}
\end{aligned}
\end{equation}
with ``body'' yet to be defined. For non-covalent interactions, the ``body'' is typically a monomer, such as a single water molecule in q-AQUA water potential.\cite{q_AQUA} In the many-body PIP approach, this expansion is applied to large molecules and the ``body'' is an atom. For hydrocarbons, this many-body expansion can be explicitly written as
\begin{equation}
\begin{aligned}
    V = & \sum_iV^{(1b)}_{\text{C}_i} + \sum_iV^{(1b)}_{\text{H}_i} + \sum_{i,j}V^{(2b)}_{\text{C}_i\text{C}_j} + \sum_{i,j}V^{(2b)}_{\text{C}_i\text{H}_j} + \sum_{i,j}V^{(2b)}_{\text{H}_i\text{H}_j} + \\
    & \sum_{i,j,k}V^{(3b)}_{\text{C}_i\text{C}_j\text{C}_k} + \sum_{i,j,k}V^{(3b)}_{\text{C}_i\text{C}_j\text{H}_k} + \sum_{i,j,k}V^{(3b)}_{\text{H}_i\text{H}_j\text{C}_k} + \sum_{i,j,k}V^{(3b)}_{\text{H}_i\text{H}_j\text{H}_k} + \\
    & \sum_{i,j,k,l} V^{(4b)}_{\text{C}_i\text{C}_j\text{C}_k\text{C}_l} + \sum_{i,j,k,l}V^{(4b)}_{\text{C}_i\text{C}_j\text{C}_k\text{H}_l} + \sum_{i,j,k,l}V^{(4b)}_{\text{C}_i\text{C}_j\text{H}_k\text{H}_l} + \sum_{i,j,k,l}V^{(4b)}_{\text{H}_i\text{H}_j\text{H}_k\text{C}_l} + \sum_{i,j,k,l}V^{(4b)}_{\text{H}_i\text{H}_j\text{H}_k\text{H}_l}
\end{aligned}
\label{eq:mbHC}
\end{equation}

The 1-body term is simply the energy of a single carbon or hydrogen atom, obtainable from electronic structure calculations at the same level of theory as the calculation for the total energy $V$. Each higher-body term in Eq. \ref{eq:mbHC} can be expressed as a linear combination in PIP basis with unknown coefficients to be determined in a fit. Take $V^{(3b)}_{\text{C}_i\text{C}_j\text{H}_k}$ as an example,
\begin{equation}
    V^{(3b)}_{\text{C}_i\text{C}_j\text{H}_k} = \sum_m c^{(\text{CCH})}_m p^{(\ce{A2B})}_m(y_{\text{C}_i\text{C}_j}, y_{\text{C}_i\text{H}_k}, y_{\text{C}_j\text{H}_k}),
\end{equation}
where the superscript ``$(\text{CCH})$'' means this set of coefficients, to be determined during training, is for the CCH 3-body term, and ``(\ce{A2B})'' represents the permutational symmetry of this 3-body term (that is, the 2 carbons can be permuted with each other, but C and H cannot be permuted). The $y_{ij}$'s are Morse-like variables of the internuclear distances $y_{ij}=e^{(-r_{ij}/\lambda)}$, where $\lambda$ is the range (hyper)parameter.  Other terms in Eq. \ref{eq:mbHC} can be expressed in a similar way using PIPs as basis. Note that all PIPs used in the MB-PES are ``purified'', i.e., the PIP is zero when any of the Morse-variable is zero.

The total number of 4-body interactions in Eq. \ref{eq:mbHC} is of the order $\mathcal{O}(N^4)$, where $N$ is the total number of atoms in a molecule. This imposes substantial computations if $N$ is large. Therefore, distance-based cutoffs are applied for $k$-body ($k=2,3,4$) terms: if the maximum internuclear distance within a $k$-mer, $r_\text{max}$, exceeds the cutoff distance, that $k$-mer is excluded from the summation. This greatly reduces the number of terms to evaluate in Eq. \ref{eq:mbHC}.
To ensure the potential is smooth and differentiable, a switching function of $r_\text{max}$ is applied:
\begin{equation}
    s = 
    \begin{cases}
    1 & r_\text{max} < r_i, \\[6pt]
    1 - 10 \left(\frac{r_\text{max}-r_i}{r_f-r_i} \right)^3 + 15\left(\frac{r_\text{max}-r_i}{r_f-r_i} \right)^4 - 6\left(\frac{r_\text{max}-r_i}{r_f-r_i} \right)^5 & r_i < r_\text{max} < r_f, \\[6pt]
    0 & r_\text{max} > r_f,
    \end{cases}
\end{equation}
where $r_i$ and $r_f$ are inner and outer cutoffs which can vary for different $k$-mers.

\subsection{Training and Test Data Sets}
\begin{figure}[H]
\includegraphics[width=0.7\textwidth]{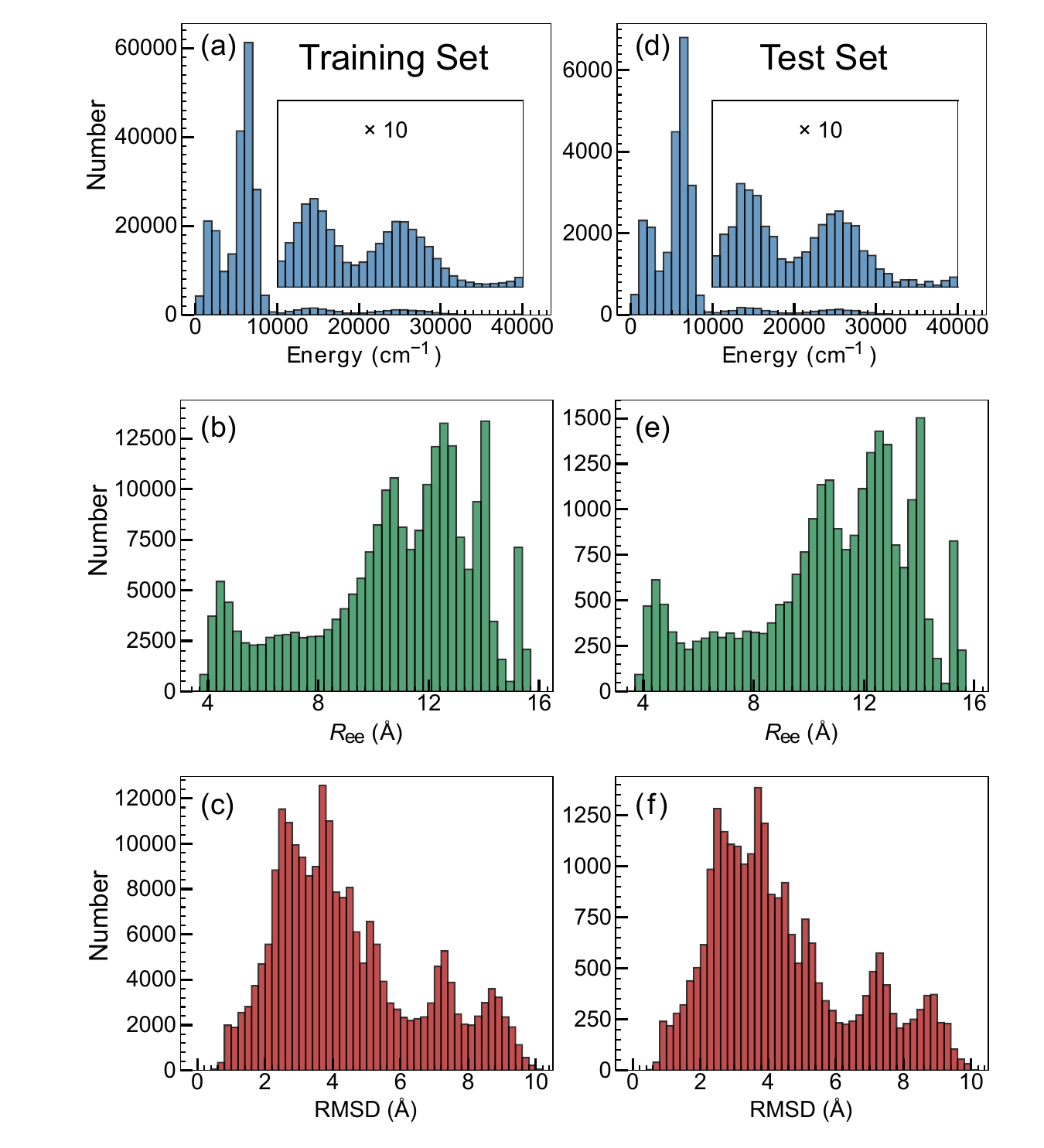}
\caption{Distribution of potential energies relative to the global minimum of \ce{C14H30}, the end-to-end carbon distance ($R_{\text{ee}}$), and the root-mean-square deviation (RMSD) with respect to the global-minimum structure for both the training (a-c) and test sets (d-f) respectively.}
\label{fig2}
\end{figure}

The original dataset used in this work was adopted from a recent study,\cite{qu2024hydrocarbon} in which a total of 272,532 configurations were sampled from molecular dynamics trajectories at various temperatures. For each configuration, single-point energies were computed at the B3LYP/cc-pVDZ level of theory. We generated a subset of these configurations through selecting structures with the DFT energies
within 40,000 cm$^{-1}$ above the global minimum. In addition, 253 configurations were excluded due to an incorrect ordering of carbon atoms in the original dataset. The resulting dataset comprises 247,211 configurations and was randomly partitioned into training and test sets using a 90:10 split. This random partition was intended to generate sets with similar data distributions. This is illustrated in Fig. \ref{fig2}, which shows the distributions of potential energies relative to the global minimum, the end-to-end carbon distance ($R_{\text{ee}}$), and the root-mean-square deviation (RMSD) with respect to the global-minimum structure for both the training and test sets. As expected, the two subsets exhibit nearly identical distributions in terms of both energetic and structural descriptors. The majority of configurations lie below 10,000 cm$^{-1}$ above the global minimum, while a substantial fraction spans a broader energy range between 10,000 and 40,000 cm$^{-1}$. In addition, a significant number of configurations display small end-to-end distances, corresponding to highly twisted molecular configurations.

\subsection{Training details}
\subsubsection{Details for MB-PIPNet}
In the \ce{C14H30} system, monomers are categorized as \ce{CH3} and \ce{CH2} groups, exhibiting \ce{A3B} and \ce{A2B} permutational symmetries respectively. Consequently, the two-body interactions are characterized by the \ce{A6B2}, \ce{A5B2},and \ce{A4B2} symmetry classes. Since the expansion size of PIPs varies with symmetry, we adjusted the polynomial degrees and truncated the basis sets to maintain a consistent number of terms across all one-body and two-body interactions. We also adjust the decay parameter $\lambda$ of PIPs based on a systematic parameter scan. These parameters are summarized in Table \ref{tab:order}.

\begin{table}[htbp!]
    \centering
    \small
    \caption {Polynomial orders and basis set sizes for PIPs of different symmetry classes.}
    \label{tab:order}
    
    \begin{threeparttable}[htbp!]
    
        \begin{tabular*}{0.85\columnwidth}{@{\extracolsep{\fill}}ccccc}
        \toprule
        Symmetry & $\lambda$/bohr & Poly. order &  Full basis size & Truncated basis size \\
        \hline
        \ce{A2B}  &  1.8 &  6 & 49 & 45 \\
        \ce{A3B}  &  1.8 &  4 & 50 & 45 \\
        \ce{A4B2}  &  1.5 &  5 & 608 & 200 \\
        \ce{A5B2}  &  1.5 &  4 & 217 & 200 \\
        \ce{A6B2}  &  1.5 &  4 & 227 & 200 \\
        \bottomrule
        \end{tabular*}
        
    \end{threeparttable}
\end{table}


In this work, two independent neural networks were established to compute the energy contributions of the methyl and methylene monomers. The input vector for each network corresponds to the descriptor $\boldsymbol{D}_i$ defined in Eq. \ref{eq:dch3} or Eq. \ref{eq:dch2}. Prior to training, both the input feature vectors and the target energy values were linearly scaled to the range of $[-1, 1]$ using min-max normalization.

To facilitate code reusability, an identical network architecture—a fully connected 445-15-15-1 topology—was adopted for both monomer types, although distinct architectures could theoretically be employed. The networks utilize the hyperbolic tangent (tanh) activation function and were optimized using the Levenberg-Marquardt (LM) algorithm\cite{more2006levenberg} for a maximum of 2000 epochs, with an early stopping mechanism implemented to prevent overfitting. For this work, no cutoff radius was applied when constructing environmental descriptors for \ce{CH2} and \ce{CH3} units. Actually and as demonstrated in our previous work\cite{yu2025extending}, a reasonable cutoff distance can be applied for large systems such as liquid water.

\subsubsection{Details for MB-PES}
The final fitting basis of MB-PES is the union of 2-body, 3-body, and 4-body bases, each could have different Morse range parameters and distance-based cutoffs. In principle, these hyperparameters can also be optimized during training or through parameter scan similar to the case of MB-PIPNet. Here, similar to our previous work,\cite{qu2024hydrocarbon} we simply set their values according to the general expectation that the $k$-body interactions fall off as $k$ increases. Table \ref{tab:aPIP} shows the value of these hyperparameters as well as the number of unknown linear coefficients for each of the $k$-body terms. The total number of unknown coefficients is 736, and these are determined all at once in a unified linear least-squares fit to the total energy of the molecule (with 1-body energy subtracted). 

\begin{table}[H]
    \centering
    \small
    \caption {Hyperparameters, polynomial order, and the number of linear coefficients in MB-PES.}
    \label{tab:aPIP}
    
    \begin{threeparttable}[htbp!]
    
        \begin{tabular*}{0.75\columnwidth}{@{\extracolsep{\fill}}cccccc}
        \toprule
        Term & $\lambda$/bohr & $r_i$/\AA & $r_f$/\AA & Poly. order & No. of coefs \\
        \hline
          CC 2-body & 2.5 & 9 & 10 & 10 &  10 \\
          CH 2-body & 2.5 & 9 & 10 & 10 &  10 \\
          HH 2-body & 2.5 & 9 & 10 & 10 &  10 \\
         CCC 3-body & 2.5 & 7 &  8 &  8 &  32 \\
         CCH 3-body & 1.8 & 7 &  8 &  8 &  78 \\
         CHH 3-body & 1.8 & 7 &  8 &  8 &  78 \\
         HHH 3-body & 1.8 & 7 &  8 &  8 &  32 \\
        CCCC 4-body & 1.2 & 5 &  6 &  6 &  40 \\
        CCCH 4-body & 1.2 & 5 &  6 &  6 & 115 \\
        CCHH 4-body & 1.2 & 5 &  6 &  6 & 174 \\
        CHHH 4-body & 1.2 & 5 &  6 &  6 & 115 \\
        HHHH 4-body & 1.2 & 5 &  6 &  6 &  40 \\
        \bottomrule
        \end{tabular*}
        
    \end{threeparttable}
\end{table}

\subsubsection{Details for DeepMD}
We also trained a neural network potential for the n-tetradecane system using the DeePMD-kit package.\cite{wang2018deepmd,DeepMD} Specifically, we employed the Deep Potential Smooth Edition (DeepPot-SE) scheme with the se\_e2\_a descriptor to capture the local atomic environments. The cutoff radius was set to 6.0 \AA, featuring a smoothing parameter of 0.5 \AA. The architecture of the embedding network consisted of three hidden layers with [100, 100, 100] neurons each, and the size of the embedding submatrix was specified as 16. The fitting network was composed of three hidden layers with [240, 240, 240] neurons. The training was conducted for a total of 200,000 steps using a batch size of 16. An exponentially decaying learning rate was applied, starting from $1.0 \times 10^{-5}$ and decreasing to a final value of $1.0 \times 10^{-8}$, with a decay interval of 1,500 steps. Validation performance was monitored and recorded every 1,000 steps throughout the training procedure.

\subsection{Molecular dynamics}

Molecular dynamics (MD) simulations were conducted to validate the stability and robustness of the developed MB-PIPNet PES. These simulations were run using our in-house quasi-classical trajectory (QCT) codes. A total of 100 microcanonical (NVE) ensemble trajectories were propagated, each with a total energy of 5000 cm$^{-1}$. All trajectories were initiated from the global minimum geometry, with the total energy randomly distributed among the kinetic energies of the atoms and the total angular momentum adjusted to zero. The simulations were run for 10,000 steps with a time step of 15 a.u. ($\sim$0.36 fs). The vibrational power spectrum was calculated using the atomic velocities recorded along MD trajectories and the Wiener-Khinchin theorem\cite{Wiener1930,Khinchin1934,Allen2017} is employed to obtain mass-weighted velocity autocorrelation function (VACF) for each of the 100 independent trajectories. The final power spectrum $I(\omega)$ was generated by applying a Fourier transform to the averaged VACFs with Hanning window function $W(t)$, such that
\begin{equation}
    I(\omega)=\left|\int_{-\infty}^{+\infty}dt\left[\bar{C_m}(t) \cdot W(t)\right]e^{-i\omega t}\right|.
\end{equation}
and 
\begin{equation}
    W(t)=0.5-0.5 \cos \left(\frac{2\pi t}{M-1}\right), 0 \leq  t \leq M-1,
\end{equation}
where $M$ is the total number of time steps in the VACF.




\section{Results}
Using the datasets and training protocols described above, we constructed three MLPs for \ce{C14H30} based on DeepMD, MB-PES, and MB-PIPNet. The corresponding training and test accuracies are summarized in Table \ref{tab:RMSE}. As shown, the MB-PIPNet model substantially outperforms the DeepMD potential for both the training and test datasets. A similar trend was observed in our previous studies of non-covalent molecular systems,\cite{yu2025extending} indicating the robustness of this behavior across different classes of molecular systems.

The improved performance of MB-PIPNet can be attributed to the use of mathematically compact PIPs to represent monomeric structures and their intermolecular interactions, leading to a more efficient and expressive structural description. When PIPs are further employed within an atomic many-body expansion of the total energy, the resulting MB-PES model achieves a slightly better test RMSE of 12.5 meV, compared to 16.0 meV for MB-PIPNet.

\begin{table}[H]
    \centering
    \small
    \caption {Model accuracies (in meV) on the \ce{C14H30} training and test dataset for DeePMD, MB-PES, and MB-PIPNet.}
    \label{tab:RMSE}
    \begin{threeparttable}[htbp!]
    
        \begin{tabular*}{0.75\columnwidth}{@{\extracolsep{\fill}}lcccc}
        \toprule
        Model & training RMSE & training MAE &  test RMSE & test MAE \\
        \hline
        \ce{DeePMD}  & 65.1 & 41.0 & 65.7 & 40.9 \\
        \ce{MB-PIPNet}  &  12.1 &  7.72 & 16.0 & 8.67 \\
        \ce{MB-PES}  &  8.80 &  4.59 & 12.5 & 4.84 \\
        \bottomrule
        \end{tabular*}
        
    \end{threeparttable}
\end{table}

\begin{figure}[H]
\includegraphics[width=0.6\textwidth]{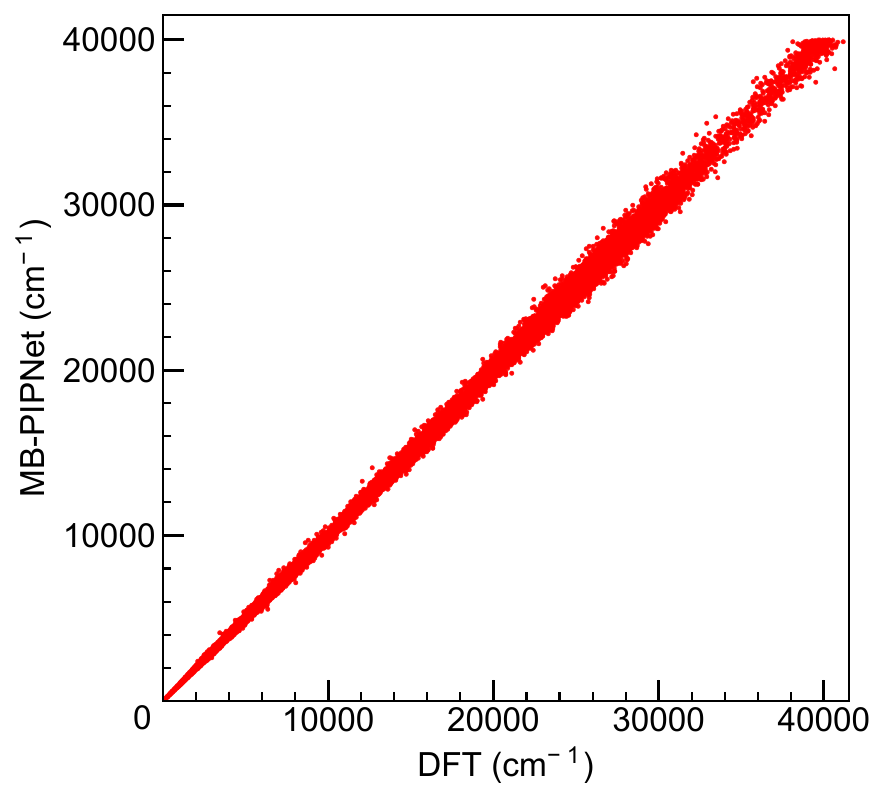}
\caption{Correlation plots of the energies for \ce{C14H30} from the MB-PIPNet PES and direct DFT B3LYP calculations.}
\label{fig:ee}
\end{figure}

\begin{figure}[H]
\includegraphics[width=0.6\textwidth]{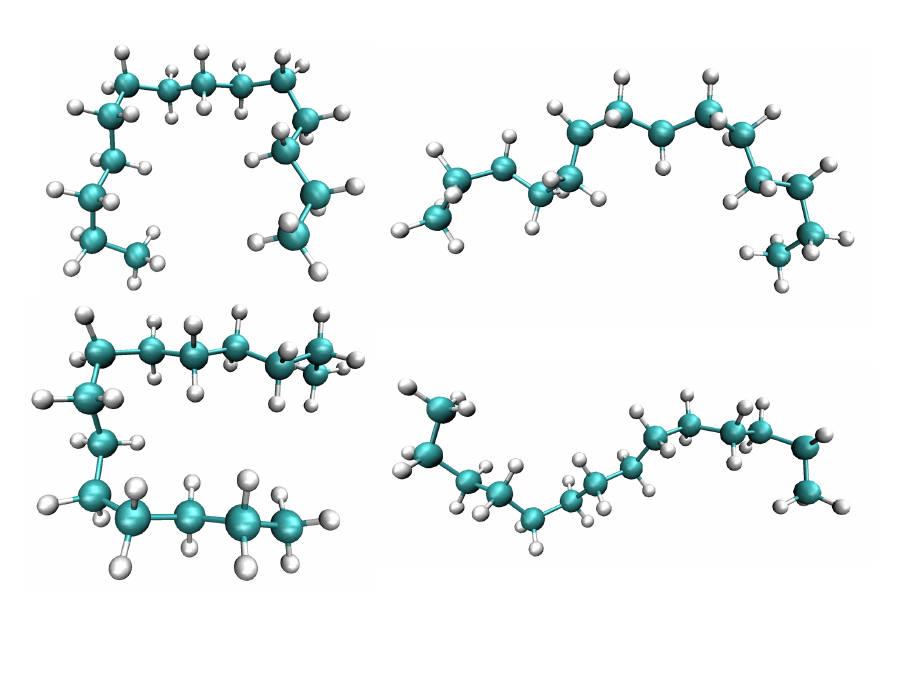}
\caption{Four selected structures from the data set}
\label{fig:struct}
\end{figure}

Figure \ref{fig:ee} shows the correlation between energies predicted by MB-PIPNet and the reference B3LYP values. Overall, excellent agreement is achieved across the dataset, with particularly high accuracy in the low-energy region (below 10,000 cm$^{-1}$). As the energy increases toward the upper end of the sampled range (up to 40,000 cm$^{-1}$), a gradual deterioration in fitting accuracy is observed, consistent with the increased structural distortion present in these configurations. Representative molecular structures from the dataset are illustrated in Figure \ref{fig:struct}, including highly twisted, high-energy conformations of \ce{C14H30}. These examples demonstrate that the MB-PIPNet model is capable of accurately describing both low-energy near-equilibrium structures and significantly distorted configurations.
 
\begin{figure}[H]
\includegraphics[width=0.6\textwidth]{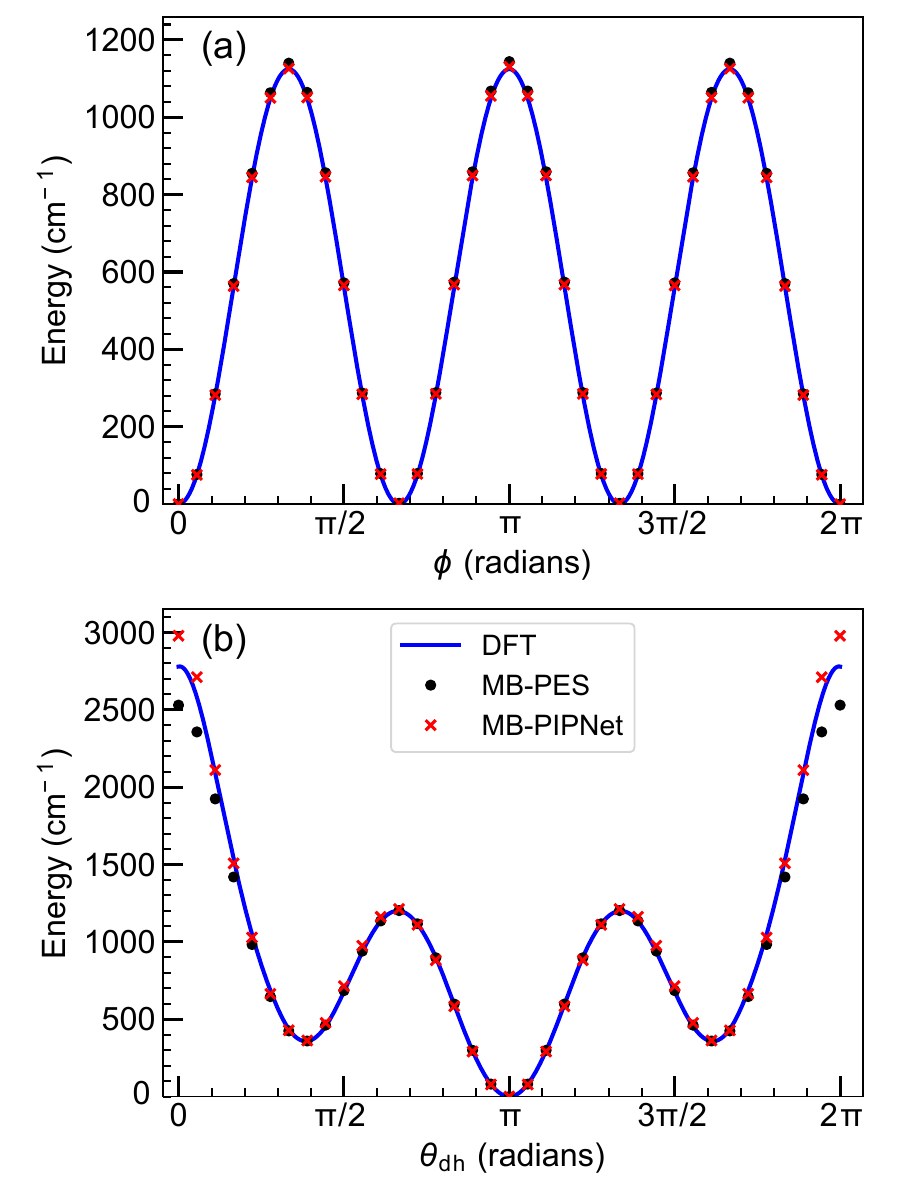}
\caption{Torsional potential energy curves for (a) methyl rotation about the C13–C14 bond and (b) ethyl rotation about the C12–C13 bond  of the global-minimum structure of \ce{C14H30}. Carbon atoms are labeled sequentially as C1–C14 along the alkane chain, starting from one terminal end.}
\label{fig:ts}
\end{figure}

We next assess the accuracy of the MB-PIPNet potential for torsional motions about C–C bonds in the global-minimum structure of \ce{C14H30} (Fig. \ref{fig1}a). The upper panel of Fig. \ref{fig:ts} presents the potential energy curves associated with methyl rotation about the C13–C14 bond, as predicted by MB-PES and MB-PIPNet, together with the reference B3LYP results. Both MLPs accurately reproduce the B3LYP torsional profile, indicating that the essential features of the methyl rotation are well captured. The performance of MB-PIPNet is further examined for ethyl rotation about the C12–C13 bond, shown in the lower panel of Fig. \ref{fig:ts}. In this case, the rotation angle is defined as the dihedral angle between the plane formed by atoms C11–C12–C13 and that formed by atoms C12–C13–C14. The global-minimum structure corresponds to a dihedral angle of $\pi$. As shown, MB-PIPNet exhibits excellent agreement with the B3LYP reference data over a wide angular range, particularly for dihedral angles between 0.5 and 5.7 radians. This region encompasses the two local minima (at approximately 1.1 and 5.05 radian), the corresponding transition states (around 2.05 and 4.1 radian), and the global minimum. In the higher-energy regions, corresponding to dihedral angles smaller than 0.5 radian or larger than 5.7 radian, the MB-PIPNet potential demonstrates slightly improved agreement with the B3LYP reference compared to MB-PES. This behavior is notable given that MB-PES achieves a lower overall training RMSE than MB-PIPNet, as discussed above, suggesting that the monomer-based representation employed in MB-PIPNet provides enhanced robustness for describing highly twisted conformations. Finally, it should be noted that for both torsional scans, all other internal coordinates were held fixed, with only the specified dihedral angle varied. Consequently, the resulting potential energy curves correspond to rigid torsional rotations. Allowing full structural relaxation at each dihedral angle would be expected to further lower the potential energy curves.

\begin{figure}[H]
\includegraphics[width=0.6\textwidth]{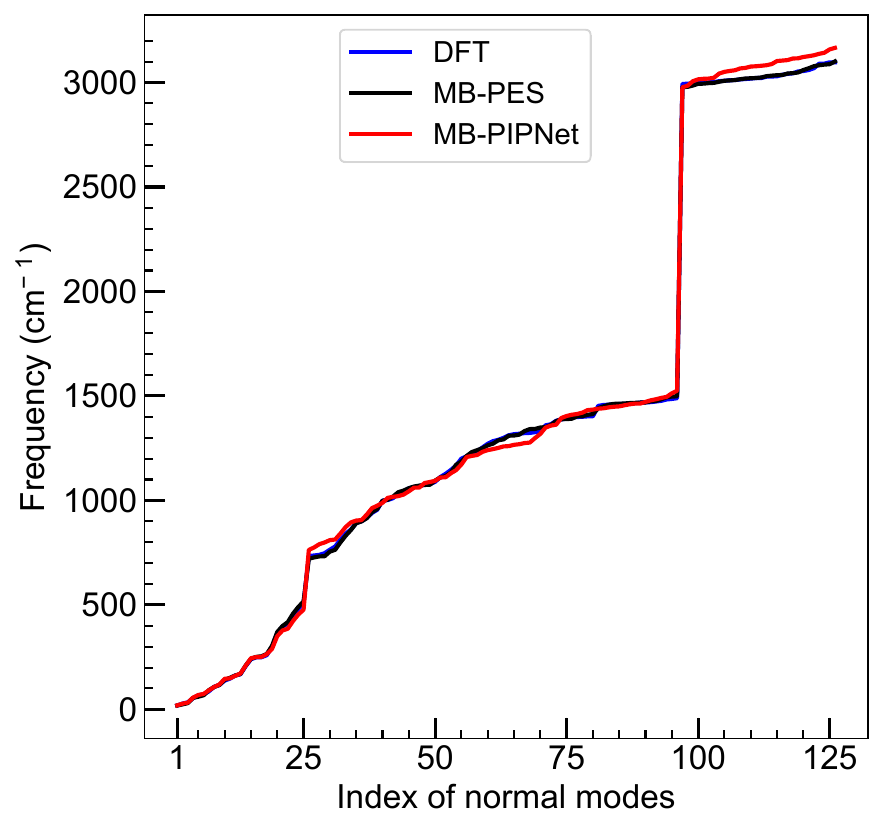}
\caption{Comparisons between the B3LYP reference harmonic frequencies and the corresponding values calculated with MB-PES and MB-PIPNet for the global minimum structure of the alkane \ce{C14H30}.}
\label{fig:freq}
\end{figure}

The MB-PIPNet potential is further evaluated through vibrational analyses of \ce{C14H30}. Starting from the global-minimum structure, standard harmonic normal-mode analysis was performed. Figure \ref{fig:freq} compares the harmonic vibrational frequencies obtained from MB-PES and MB-PIPNet with the reference B3LYP results. As shown, MB-PES exhibits excellent agreement with the DFT frequencies across nearly all 126 normal modes. The MB-PIPNet potential accurately reproduces the harmonic frequencies of low- and mid-frequency vibrational modes below approximately 1600 cm$^{-1}$. In the high-frequency C–H stretching region around 3000 cm$^{-1}$, MB-PIPNet captures the overall frequency pattern but shows a modest overestimation relative to the B3LYP reference. This deviation is likely associated with the limited set of PIP-based descriptors employed in the current monomeric neural-network representations. Investigations on careful selection of PIP bases to better capture both intramolecular distortions and environmental effects are subject to our future study. Nevertheless, the overall agreement observed here demonstrates that MB-PIPNet provides a flexible and reliable representation of the multidimensional potential energy landscape of linear alkanes, enabling accurate predictions of vibrational properties beyond simple structural and energetic benchmarks.

\begin{figure}[H]
\includegraphics[width=0.6\textwidth]{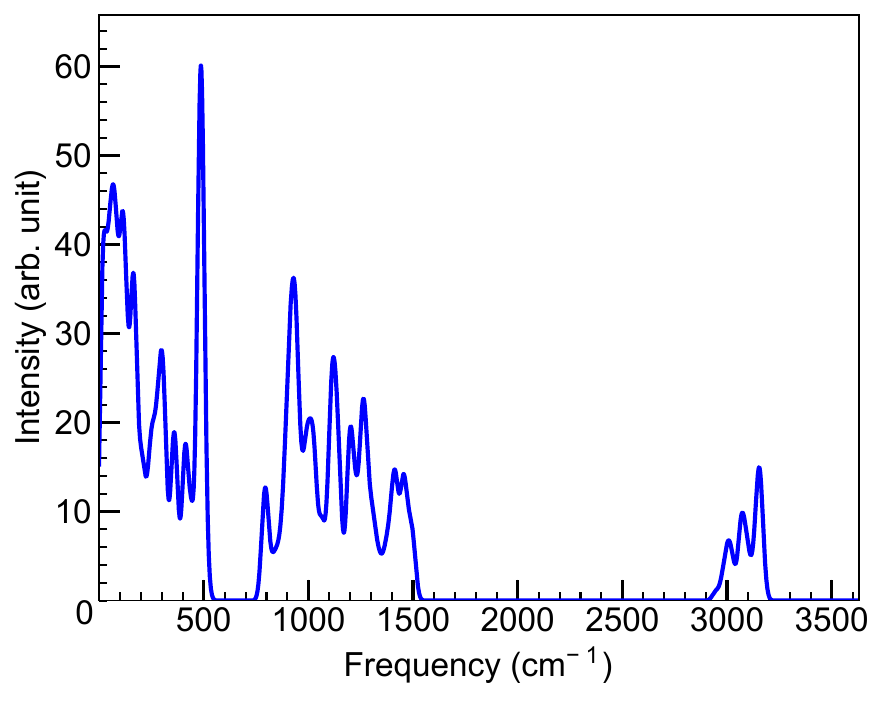}
\caption{Power spectrum of \ce{C14H30} using NVE-MD trajectories run with total energy of 5000 cm$^{-1}$.}
\label{fig:spec}
\end{figure}

Microcanonical (NVE) molecular dynamics simulations were also performed for linear \ce{C14H30}, and vibrational power spectra were obtained by Fourier transforming the velocity autocorrelation function. The resulting spectrum is shown in Fig. \ref{fig:spec}. In the high-frequency region around 3000 cm$^{-1}$, the MD-based power spectrum exhibits clear signatures associated with C–H stretching vibrations, consistent with the features identified in the harmonic normal-mode analysis. In the lower-frequency region below approximately 1600 cm$^{-1}$, the power spectrum displays a series of broad bands with multiple peaks, reflecting the complex collective motions involving C–C stretching, C–C bending, and C–H bending modes. These spectral characteristics are again in qualitative agreement with the harmonic frequency analysis discussed above. To obtain quantitatively accurate vibrational spectra for \ce{C14H30} and enable direct comparison with experimental measurements, more advanced approaches, such as path-integral molecular dynamics or fully quantum vibrational methods\cite{jpcac14h30} would be required to account for anharmonicity and nuclear quantum effects. Nevertheless, the ability of the MB-PIPNet potential to generate physically reasonable vibrational features and to stably sample relevant molecular configurations in molecular dynamics simulations demonstrates its accuracy and robustness for dynamical applications.

The final assessment of the MB-PIPNet potential concerns its computational efficiency. In our previous work,\cite{yu2025extending} we demonstrated that the MB-PIPNet model for liquid water achieves a computational cost comparable to that of conventional classical force fields such as q-TIP4P/F,\cite{q-TIP4P} while substantially outperforming atomistic MLPs such as DeepMD.\cite{DeepMD} Here, we extend this analysis to covalent systems by comparing the computational performance of MB-PIPNet, DeepMD, and MB-PES for linear alkanes.

\begin{table}[htbp!]
    \centering
    \caption{Computational time for calculations of energies and gradients of 100,000 geometries for \ce{C14H30} using different methods. All timing tests were performed using a single CPU core of the AMD EPYC 7002 processors.}
    \label{tab:ermse}
    
    \begin{threeparttable}[htbp!]
    
        \begin{tabular*}{0.75\columnwidth}{@{\extracolsep{\fill}}lcc}
        \toprule
        PES & energies/s & energies and gradients/s  \\
        \hline
        \ce{DeePMD}  & - & 1792  \\
        \ce{MB-PIPNet}  &  49 &  240  \\
        \ce{MB-PES}  &  295 &  1248  \\

        \bottomrule
        \end{tabular*}

    \end{threeparttable}
\end{table}

Specifically, the computational costs for evaluating energies and forces for 100,000 molecular geometries were measured using a single CPU core of the AMD EPYC 7002 processors. The timing results are summarized in Table \ref{tab:ermse}. As expected, the MB-PIPNet model significantly outperforms both DeepMD and MB-PES. In particular, MB-PIPNet requires only 240 s to compute both energies and gradients for 100,000 geometries, compared to 1792 s for DeepMD. In addition to this substantial speedup, MB-PIPNet also delivers markedly improved accuracy relative to DeepMD, as summarized in Table \ref{tab:RMSE} for both training and test datasets.

In comparison, the MB-PES model incurs approximately 6-8 times increase in computational cost relative to MB-PIPNet for single-point energy evaluations or MD steps involving both energies and forces. Consequently, MB-PIPNet achieves a favorable balance between accuracy and efficiency, providing model performance comparable to MB-PES while maintaining significantly lower computational cost. These results highlight the potential of the MB-PIPNet framework for large-scale quantum or classical simulations of complex molecular systems, encompassing both noncovalent and covalently bonded molecules.

\section{Conclusions}
In this work, we demonstrate a generalization of the MB-PIPNet framework\cite{yu2025extending} for constructing ML potentials for complex covalently bonded molecular systems. The proposed approach is built upon a chemically motivated decomposition of molecules into monomeric units or fragments, which allows the total potential energy to be represented as a sum of effective monomer-level contributions. PIPs are employed to represent both the internal structures of the monomeric units and their local chemical environments. These PIP-based descriptors are then combined with neural networks to accurately predict monomeric energies in a compact and physically informed manner.

The performance of this generalized MB-PIPNet framework has been systematically demonstrated using linear alkanes as a representative covalent system. The resulting MB-PIPNet potential achieves accuracy comparable to that of the atomic many-body expansion approach MB-PES, while substantially outperforming conventional invariant atomistic ML potentials such as DeePMD. The reliability of MB-PIPNet is further verified through a series of property benchmarks, including torsional potential energy curves, harmonic vibrational frequencies, and vibrational power spectra obtained from molecular dynamics simulations. A key advantage of MB-PIPNet lies in its computational efficiency. For combined energy and force evaluations, the present implementation is more than five times faster than both MB-PES and DeePMD, while maintaining comparable accuracy. This favorable balance between accuracy and efficiency makes MB-PIPNet particularly attractive for large-scale molecular simulations that are computationally prohibitive for conventional many-body or atomistic machine-learning approaches.
 
While most existing machine-learning potentials for covalent systems rely on Behler–Parrinello–type atomistic decompositions, MB-PIPNet provides an alternative route that is both chemically intuitive and computationally advantageous. It should be emphasized that the successful application of MB-PIPNet to general molecular systems requires a physically reasonable fragmentation scheme. The fragmentation idea has been extensively explored in electronic structure theory, where many-body interactions are efficiently treated through fragmentation approaches,\cite{Collins2005approximate,zhang2003molecular,fragcollins,fragzhang,fragli} such as the work of Collins and co-workers.\cite{fragcollins} More recently, Paesani and co-workers reported a general many-body expansion framework combined with fragmentation strategies for accurate machine-learning potentials of hydrocarbon systems.\cite{BullVulpe2023} Inspired by these developments, the present work suggests that MB-PIPNet can be naturally extended to more complex molecular systems containing C, H, O, and N atoms, as well as condensed-phase systems such as alcohols and related organic liquids.

Finally, we comment on the transferability of the MB-PIPNet framework. By learning effective monomeric energy contributions, MB-PIPNet shares conceptual similarities with atomistic machine-learning potentials and, in principle, enables transferability across system sizes.\cite{jiang2025review} For example, a model trained on shorter alkanes may be applied to longer chains such as \ce{C30H62}. The transferability of MB-PIPNet could be further enhanced by incorporating equivariant machine-learning architectures,\cite{MEGNet, MACE,MACE-MP-0} which will be explored in future work. Overall, we expect that the present study provides a valuable addition to modern machine-learning methodologies for molecular potential development and advances the balance among accuracy, computational efficiency, and chemical interpretability for simulations of complex molecular systems.

\section{Data Availability Statement}
The data generated and used in this study are available upon request to the authors.

\begin{acknowledgement}
Q.Y. and D.H.Z. acknowledge the support from National Natural Science Foundation of China (grant no. 22473030, 22288201, and 22533002). Q.Y. also acknowledge the funding from AI for Science Program, Shanghai Municipal Commission of Economy and Informatization (2025-GZL-RGZN-BTBX-01004).
\end{acknowledgement}

\section{Competing interests}
The authors declare no competing interests.


\bibliography{refs}

\providecommand{\latin}[1]{#1}
\makeatletter
\providecommand{\doi}
  {\begingroup\let\do\@makeother\dospecials
  \catcode`\{=1 \catcode`\}=2 \doi@aux}
\providecommand{\doi@aux}[1]{\endgroup\texttt{#1}}
\makeatother
\providecommand*\mcitethebibliography{\thebibliography}
\csname @ifundefined\endcsname{endmcitethebibliography}
  {\let\endmcitethebibliography\endthebibliography}{}
\begin{mcitethebibliography}{73}
\providecommand*\natexlab[1]{#1}
\providecommand*\mciteSetBstSublistMode[1]{}
\providecommand*\mciteSetBstMaxWidthForm[2]{}
\providecommand*\mciteBstWouldAddEndPuncttrue
  {\def\EndOfBibitem{\unskip.}}
\providecommand*\mciteBstWouldAddEndPunctfalse
  {\let\EndOfBibitem\relax}
\providecommand*\mciteSetBstMidEndSepPunct[3]{}
\providecommand*\mciteSetBstSublistLabelBeginEnd[3]{}
\providecommand*\EndOfBibitem{}
\mciteSetBstSublistMode{f}
\mciteSetBstMaxWidthForm{subitem}{(\alph{mcitesubitemcount})}
\mciteSetBstSublistLabelBeginEnd
  {\mcitemaxwidthsubitemform\space}
  {\relax}
  {\relax}

\bibitem[Gkeka \latin{et~al.}(2020)Gkeka, Stoltz, Barati~Farimani, Belkacemi,
  Ceriotti, Chodera, Dinner, Ferguson, Maillet, Minoux, \latin{et~al.}
  others]{gkeka2020machine}
Gkeka,~P.; Stoltz,~G.; Barati~Farimani,~A.; Belkacemi,~Z.; Ceriotti,~M.;
  Chodera,~J.~D.; Dinner,~A.~R.; Ferguson,~A.~L.; Maillet,~J.-B.; Minoux,~H.;
  others Machine learning force fields and coarse-grained variables in
  molecular dynamics: application to materials and biological systems. \emph{J.
  Chem. Theory Comput.} \textbf{2020}, \emph{16}, 4757--4775\relax
\mciteBstWouldAddEndPuncttrue
\mciteSetBstMidEndSepPunct{\mcitedefaultmidpunct}
{\mcitedefaultendpunct}{\mcitedefaultseppunct}\relax
\EndOfBibitem
\bibitem[Deringer \latin{et~al.}(2019)Deringer, Caro, and
  Cs{\'a}nyi]{deringer2019machine}
Deringer,~V.~L.; Caro,~M.~A.; Cs{\'a}nyi,~G. Machine learning interatomic
  potentials as emerging tools for materials science. \emph{Adv. Mat.}
  \textbf{2019}, \emph{31}, 1902765\relax
\mciteBstWouldAddEndPuncttrue
\mciteSetBstMidEndSepPunct{\mcitedefaultmidpunct}
{\mcitedefaultendpunct}{\mcitedefaultseppunct}\relax
\EndOfBibitem
\bibitem[Manzhos \latin{et~al.}(2014)Manzhos, Dawes, and Carrington]{NN-2014}
Manzhos,~S.; Dawes,~R.; Carrington,~T. Neural network-based approaches for
  building high dimensional and quantum dynamics-friendly potential energy
  surfaces. \emph{Int. J. Quantum Chem.} \textbf{2014}, \emph{115},
  1012--1020\relax
\mciteBstWouldAddEndPuncttrue
\mciteSetBstMidEndSepPunct{\mcitedefaultmidpunct}
{\mcitedefaultendpunct}{\mcitedefaultseppunct}\relax
\EndOfBibitem
\bibitem[Manzhos and Carrington~Jr(2020)Manzhos, and
  Carrington~Jr]{manzhos2020neural}
Manzhos,~S.; Carrington~Jr,~T. Neural network potential energy surfaces for
  small molecules and reactions. \emph{Chem. Rev.} \textbf{2020}, \emph{121},
  10187--10217\relax
\mciteBstWouldAddEndPuncttrue
\mciteSetBstMidEndSepPunct{\mcitedefaultmidpunct}
{\mcitedefaultendpunct}{\mcitedefaultseppunct}\relax
\EndOfBibitem
\bibitem[Meuwly(2021)]{meuwly2021machine}
Meuwly,~M. Machine learning for chemical reactions. \emph{Chem. Rev.}
  \textbf{2021}, \emph{121}, 10218--10239\relax
\mciteBstWouldAddEndPuncttrue
\mciteSetBstMidEndSepPunct{\mcitedefaultmidpunct}
{\mcitedefaultendpunct}{\mcitedefaultseppunct}\relax
\EndOfBibitem
\bibitem[Braams and Bowman(2009)Braams, and Bowman]{Braams2009}
Braams,~B.~J.; Bowman,~J.~M. Permutationally Invariant Potential Energy
  Surfaces in High Dimensionality. \emph{Int. Rev. Phys. Chem.} \textbf{2009},
  \emph{28}, 577--606\relax
\mciteBstWouldAddEndPuncttrue
\mciteSetBstMidEndSepPunct{\mcitedefaultmidpunct}
{\mcitedefaultendpunct}{\mcitedefaultseppunct}\relax
\EndOfBibitem
\bibitem[Qu \latin{et~al.}(2018)Qu, Yu, and Bowman]{ARPC2018}
Qu,~C.; Yu,~Q.; Bowman,~J.~M. Permutationally Invariant Potential Energy
  Surfaces. \emph{Annu. Rev. Phys. Chem.} \textbf{2018}, \emph{69},
  151--175\relax
\mciteBstWouldAddEndPuncttrue
\mciteSetBstMidEndSepPunct{\mcitedefaultmidpunct}
{\mcitedefaultendpunct}{\mcitedefaultseppunct}\relax
\EndOfBibitem
\bibitem[Jiang and Guo(2013)Jiang, and Guo]{PIP-NN-1}
Jiang,~B.; Guo,~H. Permutation invariant polynomial neural network approach to
  fitting potential energy surfaces. \emph{J. Chem. Phys.} \textbf{2013},
  \emph{139}, 054112\relax
\mciteBstWouldAddEndPuncttrue
\mciteSetBstMidEndSepPunct{\mcitedefaultmidpunct}
{\mcitedefaultendpunct}{\mcitedefaultseppunct}\relax
\EndOfBibitem
\bibitem[Jiang \latin{et~al.}(2016)Jiang, Li, and Guo]{Guo16}
Jiang,~B.; Li,~J.; Guo,~H. Potential energy surfaces from high fidelity fitting
  of ab initio points: The permutation invariant polynomial - neural network
  approach. \emph{Int. Rev. Phys. Chem.} \textbf{2016}, \emph{35},
  479--506\relax
\mciteBstWouldAddEndPuncttrue
\mciteSetBstMidEndSepPunct{\mcitedefaultmidpunct}
{\mcitedefaultendpunct}{\mcitedefaultseppunct}\relax
\EndOfBibitem
\bibitem[Shao \latin{et~al.}(2016)Shao, Chen, Zhao, and Zhang]{FINN1}
Shao,~K.; Chen,~J.; Zhao,~Z.; Zhang,~D.~H. Communication: Fitting potential
  energy surfaces with fundamental invariant neural network. \emph{J. Chem.
  Phys.} \textbf{2016}, \emph{145}, 071101\relax
\mciteBstWouldAddEndPuncttrue
\mciteSetBstMidEndSepPunct{\mcitedefaultmidpunct}
{\mcitedefaultendpunct}{\mcitedefaultseppunct}\relax
\EndOfBibitem
\bibitem[Fu and Zhang(2023)Fu, and Zhang]{fu2023accurate}
Fu,~B.; Zhang,~D.~H. Accurate fundamental invariant-neural network
  representation of ab initio potential energy surfaces. \emph{Natl. Sci. Rev.}
  \textbf{2023}, \emph{10}, nwad321\relax
\mciteBstWouldAddEndPuncttrue
\mciteSetBstMidEndSepPunct{\mcitedefaultmidpunct}
{\mcitedefaultendpunct}{\mcitedefaultseppunct}\relax
\EndOfBibitem
\bibitem[Behler and Parrinello(2007)Behler, and Parrinello]{BPNN}
Behler,~J.; Parrinello,~M. Generalized Neural-Network Representation of
  High-Dimensional Potential-Energy Surfaces. \emph{Phys. Rev. Lett.}
  \textbf{2007}, \emph{98}, 146401\relax
\mciteBstWouldAddEndPuncttrue
\mciteSetBstMidEndSepPunct{\mcitedefaultmidpunct}
{\mcitedefaultendpunct}{\mcitedefaultseppunct}\relax
\EndOfBibitem
\bibitem[Behler(2021)]{behler2021four}
Behler,~J. Four generations of high-dimensional neural network potentials.
  \emph{Chem. Rev.} \textbf{2021}, \emph{121}, 10037--10072\relax
\mciteBstWouldAddEndPuncttrue
\mciteSetBstMidEndSepPunct{\mcitedefaultmidpunct}
{\mcitedefaultendpunct}{\mcitedefaultseppunct}\relax
\EndOfBibitem
\bibitem[Chmiela \latin{et~al.}(2018)Chmiela, Sauceda, M{\"u}ller, and
  Tkatchenko]{chmiela2018towards}
Chmiela,~S.; Sauceda,~H.~E.; M{\"u}ller,~K.-R.; Tkatchenko,~A. Towards exact
  molecular dynamics simulations with machine-learned force fields. \emph{Nat.
  Commun.} \textbf{2018}, \emph{9}, 3887\relax
\mciteBstWouldAddEndPuncttrue
\mciteSetBstMidEndSepPunct{\mcitedefaultmidpunct}
{\mcitedefaultendpunct}{\mcitedefaultseppunct}\relax
\EndOfBibitem
\bibitem[Bart{\'o}k \latin{et~al.}(2010)Bart{\'o}k, Payne, Kondor, and
  Cs{\'a}nyi]{bartok2010gaussian}
Bart{\'o}k,~A.~P.; Payne,~M.~C.; Kondor,~R.; Cs{\'a}nyi,~G. Gaussian
  approximation potentials: The accuracy of quantum mechanics, without the
  electrons. \emph{Phys. Rev. Lett.} \textbf{2010}, \emph{104}, 136403\relax
\mciteBstWouldAddEndPuncttrue
\mciteSetBstMidEndSepPunct{\mcitedefaultmidpunct}
{\mcitedefaultendpunct}{\mcitedefaultseppunct}\relax
\EndOfBibitem
\bibitem[Uteva \latin{et~al.}(2017)Uteva, Graham, Wilkinson, and
  Wheatley]{GP-2017-1}
Uteva,~E.; Graham,~R.~S.; Wilkinson,~R.~D.; Wheatley,~R.~J. Interpolation of
  intermolecular potentials using Gaussian processes. \emph{J. Chem. Phys.}
  \textbf{2017}, \emph{147}, 161706\relax
\mciteBstWouldAddEndPuncttrue
\mciteSetBstMidEndSepPunct{\mcitedefaultmidpunct}
{\mcitedefaultendpunct}{\mcitedefaultseppunct}\relax
\EndOfBibitem
\bibitem[Sch\"{u}tt \latin{et~al.}(2018)Sch\"{u}tt, Sauceda, Kindermans,
  Tkatchenko, and M\"{u}ller]{SchNet}
Sch\"{u}tt,~K.~T.; Sauceda,~H.~E.; Kindermans,~P.-J.; Tkatchenko,~A.;
  M\"{u}ller,~K.-R. SchNet - A deep learning architecture for molecules and
  materials. \emph{J. Chem. Phys.} \textbf{2018}, \emph{148}, 241722\relax
\mciteBstWouldAddEndPuncttrue
\mciteSetBstMidEndSepPunct{\mcitedefaultmidpunct}
{\mcitedefaultendpunct}{\mcitedefaultseppunct}\relax
\EndOfBibitem
\bibitem[Unke and Meuwly(2019)Unke, and Meuwly]{PhysNet}
Unke,~O.~T.; Meuwly,~M. PhysNet: A Neural Network for Predicting Energies,
  Forces, Dipole Moments, and Partial Charges. \emph{J. Chem. Theory Comput.}
  \textbf{2019}, \emph{15}, 3678--3693\relax
\mciteBstWouldAddEndPuncttrue
\mciteSetBstMidEndSepPunct{\mcitedefaultmidpunct}
{\mcitedefaultendpunct}{\mcitedefaultseppunct}\relax
\EndOfBibitem
\bibitem[Zhang \latin{et~al.}(2018)Zhang, Han, Wang, Car, and Weinan]{DeepMD}
Zhang,~L.; Han,~J.; Wang,~H.; Car,~R.; Weinan,~E. Deep Potential Molecular
  Dynamics: A Scalable Model with the Accuracy of Quantum Mechanics.
  \emph{Phys. Rev. Lett.} \textbf{2018}, \emph{120}, 143001\relax
\mciteBstWouldAddEndPuncttrue
\mciteSetBstMidEndSepPunct{\mcitedefaultmidpunct}
{\mcitedefaultendpunct}{\mcitedefaultseppunct}\relax
\EndOfBibitem
\bibitem[Zhang \latin{et~al.}(2019)Zhang, Hu, and Jiang]{EANN}
Zhang,~Y.; Hu,~C.; Jiang,~B. Embedded atom neural network potentials: Efficient
  and accurate machine learning with a physically inspired representation.
  \emph{J. Phys. Chem. Lett.} \textbf{2019}, \emph{10}, 4962--4967\relax
\mciteBstWouldAddEndPuncttrue
\mciteSetBstMidEndSepPunct{\mcitedefaultmidpunct}
{\mcitedefaultendpunct}{\mcitedefaultseppunct}\relax
\EndOfBibitem
\bibitem[Batzner \latin{et~al.}(2022)Batzner, Musaelian, Sun, Geiger, Mailoa,
  Kornbluth, Molinari, Smidt, and Kozinsky]{NequIP}
Batzner,~S.; Musaelian,~A.; Sun,~L.; Geiger,~M.; Mailoa,~J.~P.; Kornbluth,~M.;
  Molinari,~N.; Smidt,~T.~E.; Kozinsky,~B. E(3)-equivariant graph neural
  networks for data-efficient and accurate interatomic potentials. \emph{Nat.
  Commun.} \textbf{2022}, \emph{13}, 2453\relax
\mciteBstWouldAddEndPuncttrue
\mciteSetBstMidEndSepPunct{\mcitedefaultmidpunct}
{\mcitedefaultendpunct}{\mcitedefaultseppunct}\relax
\EndOfBibitem
\bibitem[Batatia \latin{et~al.}(2022)Batatia, Kovacs, Simm, Ortner, and
  Csanyi]{MACE}
Batatia,~I.; Kovacs,~D.~P.; Simm,~G.; Ortner,~C.; Csanyi,~G. MACE: Higher Order
  Equivariant Message Passing Neural Networks for Fast and Accurate Force
  Fields. Advances in Neural Information Processing Systems. 2022; pp
  11423--11436\relax
\mciteBstWouldAddEndPuncttrue
\mciteSetBstMidEndSepPunct{\mcitedefaultmidpunct}
{\mcitedefaultendpunct}{\mcitedefaultseppunct}\relax
\EndOfBibitem
\bibitem[Musaelian \latin{et~al.}(2023)Musaelian, Batzner, Johansson, Sun,
  Owen, Kornbluth, and Kozinsky]{Allegro}
Musaelian,~A.; Batzner,~S.; Johansson,~A.; Sun,~L.; Owen,~C.~J.; Kornbluth,~M.;
  Kozinsky,~B. Learning local equivariant representations for large-scale
  atomistic dynamics. \emph{Nat. Commun.} \textbf{2023}, \emph{14}, 579\relax
\mciteBstWouldAddEndPuncttrue
\mciteSetBstMidEndSepPunct{\mcitedefaultmidpunct}
{\mcitedefaultendpunct}{\mcitedefaultseppunct}\relax
\EndOfBibitem
\bibitem[Xia \latin{et~al.}(2025)Xia, Zhang, and Jiang]{jiang2025review}
Xia,~J.; Zhang,~Y.; Jiang,~B. The evolution of machine learning potentials for
  molecules{,} reactions and materials. \emph{Chem. Soc. Rev.} \textbf{2025},
  \emph{54}, 4790--4821\relax
\mciteBstWouldAddEndPuncttrue
\mciteSetBstMidEndSepPunct{\mcitedefaultmidpunct}
{\mcitedefaultendpunct}{\mcitedefaultseppunct}\relax
\EndOfBibitem
\bibitem[Heidar-Zadeh \latin{et~al.}(2017)Heidar-Zadeh, Ayers, Verstraelen,
  Vinogradov, V\"{o}hringer-Martinez, and Bultinck]{heidar2017information}
Heidar-Zadeh,~F.; Ayers,~P.~W.; Verstraelen,~T.; Vinogradov,~I.;
  V\"{o}hringer-Martinez,~E.; Bultinck,~P. Information-theoretic approaches to
  atoms-in-molecules: Hirshfeld family of partitioning schemes. \emph{J. Phys.
  Chem. A} \textbf{2017}, \emph{122}, 4219--4245\relax
\mciteBstWouldAddEndPuncttrue
\mciteSetBstMidEndSepPunct{\mcitedefaultmidpunct}
{\mcitedefaultendpunct}{\mcitedefaultseppunct}\relax
\EndOfBibitem
\bibitem[Uhlig \latin{et~al.}(2025)Uhlig, Tovey, and
  Holm]{uhligEmergenceAccurateAtomic2025}
Uhlig,~F.; Tovey,~S.; Holm,~C. Emergence of Accurate Atomic Energies from
  Machine-Learned Noble-Gas Potentials. \emph{J. Chem. Phys.} \textbf{2025},
  \emph{162}, 181101\relax
\mciteBstWouldAddEndPuncttrue
\mciteSetBstMidEndSepPunct{\mcitedefaultmidpunct}
{\mcitedefaultendpunct}{\mcitedefaultseppunct}\relax
\EndOfBibitem
\bibitem[Konovalov \latin{et~al.}(2021)Konovalov, Symons, and Popelier]{FFLUX}
Konovalov,~A.; Symons,~B.~C.; Popelier,~P.~L. On the many-body nature of
  intramolecular forces in FFLUX and its implications. \emph{J. Comput. Chem.}
  \textbf{2021}, \emph{42}, 107--116\relax
\mciteBstWouldAddEndPuncttrue
\mciteSetBstMidEndSepPunct{\mcitedefaultmidpunct}
{\mcitedefaultendpunct}{\mcitedefaultseppunct}\relax
\EndOfBibitem
\bibitem[Bader(1985)]{QCTo1}
Bader,~R. F.~W. Atoms in molecules. \emph{Acc. Chem. Res.} \textbf{1985},
  \emph{18}, 9--15\relax
\mciteBstWouldAddEndPuncttrue
\mciteSetBstMidEndSepPunct{\mcitedefaultmidpunct}
{\mcitedefaultendpunct}{\mcitedefaultseppunct}\relax
\EndOfBibitem
\bibitem[Xie and Bowman(2010)Xie, and Bowman]{Xie10}
Xie,~Z.; Bowman,~J.~M. Permutationally invariant polynomial basis for molecular
  energy surface fitting via monomial symmetrization. \emph{J. Chem. Theory
  Comput.} \textbf{2010}, \emph{6}, 26--34\relax
\mciteBstWouldAddEndPuncttrue
\mciteSetBstMidEndSepPunct{\mcitedefaultmidpunct}
{\mcitedefaultendpunct}{\mcitedefaultseppunct}\relax
\EndOfBibitem
\bibitem[Bowman \latin{et~al.}(2011)Bowman, Cza{\'k}o, and Fu]{Bowman2011}
Bowman,~J.~M.; Cza{\'k}o,~G.; Fu,~B. High-dimensional ab initio potential
  energy surfaces for reaction dynamics calculations. \emph{Phys. Chem. Chem.
  Phys.} \textbf{2011}, \emph{13}, 8094--8111\relax
\mciteBstWouldAddEndPuncttrue
\mciteSetBstMidEndSepPunct{\mcitedefaultmidpunct}
{\mcitedefaultendpunct}{\mcitedefaultseppunct}\relax
\EndOfBibitem
\bibitem[Czakó \latin{et~al.}(2021)Czakó, Gy\H{o}ri, Papp, Tajti, and
  Tasi]{czako2021}
Czakó,~G.; Gy\H{o}ri,~T.; Papp,~D.; Tajti,~V.; Tasi,~D.~A. First-Principles
  reaction dynamics beyond six-atom systems. \emph{J. Phys. Chem. A}
  \textbf{2021}, \emph{125}, 2385--2393\relax
\mciteBstWouldAddEndPuncttrue
\mciteSetBstMidEndSepPunct{\mcitedefaultmidpunct}
{\mcitedefaultendpunct}{\mcitedefaultseppunct}\relax
\EndOfBibitem
\bibitem[Bull-Vulpe \latin{et~al.}(2023)Bull-Vulpe, Riera, Bore, and
  Paesani]{BullVulpe2023}
Bull-Vulpe,~E.~F.; Riera,~M.; Bore,~S.~L.; Paesani,~F. Data-Driven Many-Body
  Potential Energy Functions for Generic Molecules: Linear Alkanes as a
  Proof-of-Concept Application. \emph{J. Chem. Theory Comput.} \textbf{2023},
  \emph{19}, 4494--4509\relax
\mciteBstWouldAddEndPuncttrue
\mciteSetBstMidEndSepPunct{\mcitedefaultmidpunct}
{\mcitedefaultendpunct}{\mcitedefaultseppunct}\relax
\EndOfBibitem
\bibitem[Homayoon \latin{et~al.}(2015)Homayoon, Conte, Qu, and
  Bowman]{purified15c}
Homayoon,~Z.; Conte,~R.; Qu,~C.; Bowman,~J.~M. Full-dimensional, high-level ab
  initio potential energy surfaces for \ce{H2(H2O)} and \ce{H2(H2O)2} with
  application to hydrogen clathrate hydrates. \emph{J. Chem. Phys.}
  \textbf{2015}, \emph{143}, 084302\relax
\mciteBstWouldAddEndPuncttrue
\mciteSetBstMidEndSepPunct{\mcitedefaultmidpunct}
{\mcitedefaultendpunct}{\mcitedefaultseppunct}\relax
\EndOfBibitem
\bibitem[Qu \latin{et~al.}(2015)Qu, Conte, Houston, and Bowman]{Chen15}
Qu,~C.; Conte,~R.; Houston,~P.~L.; Bowman,~J.~M. ``Plug and play"
  full-dimensional ab initio potential energy and dipole moment surfaces and
  anharmonic vibrational analysis for CH4-H2O. \emph{Phys. Chem. Chem. Phys.}
  \textbf{2015}, \emph{17}, 8172--8181\relax
\mciteBstWouldAddEndPuncttrue
\mciteSetBstMidEndSepPunct{\mcitedefaultmidpunct}
{\mcitedefaultendpunct}{\mcitedefaultseppunct}\relax
\EndOfBibitem
\bibitem[Yu and Bowman(2017)Yu, and Bowman]{yujcp}
Yu,~Q.; Bowman,~J.~M. Communication: VSCF/VCI vibrational spectroscopy of
  \ce{H7O3+} and \ce{H9O4+} using high-level, many-body potential energy
  surface and dipole moment surfaces. \emph{J. Chem. Phys.} \textbf{2017},
  \emph{146}, 121102\relax
\mciteBstWouldAddEndPuncttrue
\mciteSetBstMidEndSepPunct{\mcitedefaultmidpunct}
{\mcitedefaultendpunct}{\mcitedefaultseppunct}\relax
\EndOfBibitem
\bibitem[Wang \latin{et~al.}(2009)Wang, Shepler, Braams, and Bowman]{WHBB}
Wang,~Y.~M.; Shepler,~B.~C.; Braams,~B.~J.; Bowman,~J.~M. Full-Dimensional, Ab
  Initio Potential Energy and Dipole Moment Surfaces for Water. \emph{J. Chem.
  Phys.} \textbf{2009}, \emph{131}, 054511\relax
\mciteBstWouldAddEndPuncttrue
\mciteSetBstMidEndSepPunct{\mcitedefaultmidpunct}
{\mcitedefaultendpunct}{\mcitedefaultseppunct}\relax
\EndOfBibitem
\bibitem[Reddy \latin{et~al.}(2016)Reddy, Straight, Bajaj, Huy~Pham, Riera,
  Moberg, Morales, Knight, Götz, and Paesani]{mbpoltests}
Reddy,~S.~K.; Straight,~S.~C.; Bajaj,~P.; Huy~Pham,~C.; Riera,~M.;
  Moberg,~D.~R.; Morales,~M.~A.; Knight,~C.; Götz,~A.~W.; Paesani,~F. On the
  Accuracy of the MB-pol Many-body Potential for Water: Interaction Energies,
  Vibrational Frequencies, and Classical Thermodynamic and Dynamical Properties
  from Clusters to Liquid water and Ice. \emph{J. Chem. Phys.} \textbf{2016},
  \emph{145}, 194504\relax
\mciteBstWouldAddEndPuncttrue
\mciteSetBstMidEndSepPunct{\mcitedefaultmidpunct}
{\mcitedefaultendpunct}{\mcitedefaultseppunct}\relax
\EndOfBibitem
\bibitem[Yu \latin{et~al.}(2022)Yu, Qu, Houston, Conte, Nandi, and
  Bowman]{q_AQUA}
Yu,~Q.; Qu,~C.; Houston,~P.~L.; Conte,~R.; Nandi,~A.; Bowman,~J.~M. q-AQUA: A
  many-body CCSD(T) water potential, including 4-body interactions,
  demonstrates the quantum nature of water from clusters to the liquid phase.
  \emph{J. Phys. Chem. Letts.} \textbf{2022}, \emph{13}, 5068--5074\relax
\mciteBstWouldAddEndPuncttrue
\mciteSetBstMidEndSepPunct{\mcitedefaultmidpunct}
{\mcitedefaultendpunct}{\mcitedefaultseppunct}\relax
\EndOfBibitem
\bibitem[Zhu \latin{et~al.}(2023)Zhu, Riera, Bull-Vulpe, and Paesani]{Mbpol23}
Zhu,~X.; Riera,~M.; Bull-Vulpe,~E.~F.; Paesani,~F. MB-pol(2023): Sub-chemical
  Accuracy for Water Simulations from the Gas to the Liquid Phase. \emph{J.
  Chem. Theory Comput} \textbf{2023}, \emph{19}, 3551--3556\relax
\mciteBstWouldAddEndPuncttrue
\mciteSetBstMidEndSepPunct{\mcitedefaultmidpunct}
{\mcitedefaultendpunct}{\mcitedefaultseppunct}\relax
\EndOfBibitem
\bibitem[Qu \latin{et~al.}(2023)Qu, Yu, Houston, Conte, Nandi, and
  Bowman]{qAQUApol}
Qu,~C.; Yu,~Q.; Houston,~P.~L.; Conte,~R.; Nandi,~A.; Bowman,~J.~M. Interfacing
  q-AQUA with a Polarizable Force Field: The Best of Both Worlds. \emph{J.
  Chem. Theory Comput} \textbf{2023}, \emph{19}, 3446--3459\relax
\mciteBstWouldAddEndPuncttrue
\mciteSetBstMidEndSepPunct{\mcitedefaultmidpunct}
{\mcitedefaultendpunct}{\mcitedefaultseppunct}\relax
\EndOfBibitem
\bibitem[van~der Oord \latin{et~al.}(2020)van~der Oord, Dusson, Csányi, and
  Ortner]{aPIP2020}
van~der Oord,~C.; Dusson,~G.; Csányi,~G.; Ortner,~C. Regularised atomic
  body-ordered permutation-invariant polynomials for the construction of
  interatomic potentials. \emph{Mach. Learn.: Sci. Technol.} \textbf{2020},
  \emph{1}, 015004\relax
\mciteBstWouldAddEndPuncttrue
\mciteSetBstMidEndSepPunct{\mcitedefaultmidpunct}
{\mcitedefaultendpunct}{\mcitedefaultseppunct}\relax
\EndOfBibitem
\bibitem[Fu and Zhang(2018)Fu, and Zhang]{Fu18}
Fu,~B.; Zhang,~D.~H. Ab initio potential energy surfaces and quantum dynamics
  for polyatomic bimolecular reactions. \emph{J. Chem. Theory Comput.}
  \textbf{2018}, \emph{14}, 2289--2303\relax
\mciteBstWouldAddEndPuncttrue
\mciteSetBstMidEndSepPunct{\mcitedefaultmidpunct}
{\mcitedefaultendpunct}{\mcitedefaultseppunct}\relax
\EndOfBibitem
\bibitem[Qu \latin{et~al.}(2018)Qu, Yu, Van~Hoozen, Bowman, and
  Vargas-Hern{\'a}ndez]{PIP-GP}
Qu,~C.; Yu,~Q.; Van~Hoozen,~B.~L.; Bowman,~J.~M.; Vargas-Hern{\'a}ndez,~R.~A.
  Assessing Gaussian Process Regression and Permutationally Invariant
  Polynomial Approaches To Represent High-Dimensional Potential Energy
  Surfaces. \emph{J. Chem. Theory Comput.} \textbf{2018}, \emph{14},
  3381--3396\relax
\mciteBstWouldAddEndPuncttrue
\mciteSetBstMidEndSepPunct{\mcitedefaultmidpunct}
{\mcitedefaultendpunct}{\mcitedefaultseppunct}\relax
\EndOfBibitem
\bibitem[Bowman \latin{et~al.}(2025)Bowman, Qu, Conte, Nandi, Houston, and
  Yu]{bowman2025perspective}
Bowman,~J.~M.; Qu,~C.; Conte,~R.; Nandi,~A.; Houston,~P.~L.; Yu,~Q. A
  Perspective Marking 20 Years of Using Permutationally Invariant Polynomials
  for Molecular Potentials. \emph{J. Chem. Phys.} \textbf{2025}, \emph{162},
  180901\relax
\mciteBstWouldAddEndPuncttrue
\mciteSetBstMidEndSepPunct{\mcitedefaultmidpunct}
{\mcitedefaultendpunct}{\mcitedefaultseppunct}\relax
\EndOfBibitem
\bibitem[Conte \latin{et~al.}(2015)Conte, Qu, and Bowman]{purified15a}
Conte,~R.; Qu,~C.; Bowman,~J.~M. Permutationally Invariant Fitting of
  Many-Body, Non-covalent Interactions with Application to Three-Body
  Methane–Water–Water. \emph{J. Chem. Theory Comput.} \textbf{2015},
  \emph{11}, 1631\relax
\mciteBstWouldAddEndPuncttrue
\mciteSetBstMidEndSepPunct{\mcitedefaultmidpunct}
{\mcitedefaultendpunct}{\mcitedefaultseppunct}\relax
\EndOfBibitem
\bibitem[Bull-Vulpe \latin{et~al.}(2021)Bull-Vulpe, Riera, G{\"o}tz, and
  Paesani]{MBfit}
Bull-Vulpe,~E.~F.; Riera,~M.; G{\"o}tz,~A.~W.; Paesani,~F. MB-Fit: Software
  infrastructure for data-driven many-body potential energy functions. \emph{J.
  Chem. Phys.} \textbf{2021}, \emph{155}, 124801\relax
\mciteBstWouldAddEndPuncttrue
\mciteSetBstMidEndSepPunct{\mcitedefaultmidpunct}
{\mcitedefaultendpunct}{\mcitedefaultseppunct}\relax
\EndOfBibitem
\bibitem[Gupta \latin{et~al.}(2025)Gupta, Bull-Vulpe, Agnew, Iyer, Zhu, Zhou,
  Knight, and Paesani]{MBX1.2}
Gupta,~S.; Bull-Vulpe,~E.~F.; Agnew,~H.; Iyer,~S.; Zhu,~X.; Zhou,~R.;
  Knight,~C.; Paesani,~F. MBX V1. 2: Accelerating data-driven many-body
  molecular dynamics simulations. \emph{J. Chem. Theory Comput.} \textbf{2025},
  \emph{21}, 1838--1849\relax
\mciteBstWouldAddEndPuncttrue
\mciteSetBstMidEndSepPunct{\mcitedefaultmidpunct}
{\mcitedefaultendpunct}{\mcitedefaultseppunct}\relax
\EndOfBibitem
\bibitem[Zhu \latin{et~al.}(2022)Zhu, Yang, Zeng, Fang, Jiang, Zhang, and
  Li]{zhangFIMB2022}
Zhu,~Y.-C.; Yang,~S.; Zeng,~J.-X.; Fang,~W.; Jiang,~L.; Zhang,~D.~H.; Li,~X.-Z.
  Torsional tunneling splitting in a water trimer. \emph{J. Am. Chem. Soc.}
  \textbf{2022}, \emph{144}, 21356--21362\relax
\mciteBstWouldAddEndPuncttrue
\mciteSetBstMidEndSepPunct{\mcitedefaultmidpunct}
{\mcitedefaultendpunct}{\mcitedefaultseppunct}\relax
\EndOfBibitem
\bibitem[Yu \latin{et~al.}(2025)Yu, Ma, Qu, Conte, Nandi, Pandey, Houston,
  Zhang, and Bowman]{yu2025extending}
Yu,~Q.; Ma,~R.; Qu,~C.; Conte,~R.; Nandi,~A.; Pandey,~P.; Houston,~P.~L.;
  Zhang,~D.~H.; Bowman,~J.~M. Extending atomic decomposition and many-body
  representation with a chemistry-motivated approach to machine learning
  potentials. \emph{Nat. Comput. Sci.} \textbf{2025}, \emph{5}, 418--426\relax
\mciteBstWouldAddEndPuncttrue
\mciteSetBstMidEndSepPunct{\mcitedefaultmidpunct}
{\mcitedefaultendpunct}{\mcitedefaultseppunct}\relax
\EndOfBibitem
\bibitem[Qu and Bowman(2019)Qu, and Bowman]{qu2019fragmented}
Qu,~C.; Bowman,~J.~M. A fragmented, permutationally invariant polynomial
  approach for potential energy surfaces of large molecules: Application to
  N-methyl acetamide. \emph{J. Chem. Phys.} \textbf{2019}, \emph{150},
  141101\relax
\mciteBstWouldAddEndPuncttrue
\mciteSetBstMidEndSepPunct{\mcitedefaultmidpunct}
{\mcitedefaultendpunct}{\mcitedefaultseppunct}\relax
\EndOfBibitem
\bibitem[Conte \latin{et~al.}(2020)Conte, Qu, Houston, and Bowman]{conte20}
Conte,~R.; Qu,~C.; Houston,~P.~L.; Bowman,~J.~M. Efficient Generation of
  Permutationally Invariant Potential Energy Surfaces for Large Molecules.
  \emph{J. Chem. Theory Comput.} \textbf{2020}, \emph{16}, 3264--3272\relax
\mciteBstWouldAddEndPuncttrue
\mciteSetBstMidEndSepPunct{\mcitedefaultmidpunct}
{\mcitedefaultendpunct}{\mcitedefaultseppunct}\relax
\EndOfBibitem
\bibitem[Qu \latin{et~al.}(2024)Qu, Houston, Allison, Schneider, and
  Bowman]{qu2024hydrocarbon}
Qu,~C.; Houston,~P.~L.; Allison,~T.; Schneider,~B.~I.; Bowman,~J.~M. DFT-based
  permutationally invariant polynomial potentials capture the twists and turns
  of C14H30. \emph{J. Chem. Theory Comput.} \textbf{2024}, \emph{20},
  9339--9353\relax
\mciteBstWouldAddEndPuncttrue
\mciteSetBstMidEndSepPunct{\mcitedefaultmidpunct}
{\mcitedefaultendpunct}{\mcitedefaultseppunct}\relax
\EndOfBibitem
\bibitem[Qu \latin{et~al.}(2024)Qu, Houston, Conte, and Bowman]{qu2024dynamics}
Qu,~C.; Houston,~P.~L.; Conte,~R.; Bowman,~J.~M. Dynamics calculations of the
  flexibility and vibrational spectrum of the linear alkane C14H30, based on
  machine-learned potentials. \emph{J. Phys. Chem. A} \textbf{2024},
  \emph{128}, 10713--10722\relax
\mciteBstWouldAddEndPuncttrue
\mciteSetBstMidEndSepPunct{\mcitedefaultmidpunct}
{\mcitedefaultendpunct}{\mcitedefaultseppunct}\relax
\EndOfBibitem
\bibitem[Zhang and Zhang(2003)Zhang, and Zhang]{zhang2003molecular}
Zhang,~D.~W.; Zhang,~J. Molecular fractionation with conjugate caps for full
  quantum mechanical calculation of protein--molecule interaction energy.
  \emph{J. Chem. Phys.} \textbf{2003}, \emph{119}, 3599--3605\relax
\mciteBstWouldAddEndPuncttrue
\mciteSetBstMidEndSepPunct{\mcitedefaultmidpunct}
{\mcitedefaultendpunct}{\mcitedefaultseppunct}\relax
\EndOfBibitem
\bibitem[Deev and Collins(2005)Deev, and Collins]{Collins2005approximate}
Deev,~V.; Collins,~M.~A. Approximate ab initio energies by systematic molecular
  fragmentation. \emph{J. Chem. Phys.} \textbf{2005}, \emph{122}, 154102\relax
\mciteBstWouldAddEndPuncttrue
\mciteSetBstMidEndSepPunct{\mcitedefaultmidpunct}
{\mcitedefaultendpunct}{\mcitedefaultseppunct}\relax
\EndOfBibitem
\bibitem[Collins(2007)]{collins2007molecular}
Collins,~M.~A. Molecular potential energy surfaces constructed from
  interpolation of systematic fragment surfaces. \emph{J. Chem. Phys.}
  \textbf{2007}, \emph{127}, 024104\relax
\mciteBstWouldAddEndPuncttrue
\mciteSetBstMidEndSepPunct{\mcitedefaultmidpunct}
{\mcitedefaultendpunct}{\mcitedefaultseppunct}\relax
\EndOfBibitem
\bibitem[Sahu and Gadre(2014)Sahu, and Gadre]{Gadre2014molecular}
Sahu,~N.; Gadre,~S.~R. Molecular tailoring approach: a route for ab initio
  treatment of large clusters. \emph{Acc. Chem. Res.} \textbf{2014}, \emph{47},
  2739--2747\relax
\mciteBstWouldAddEndPuncttrue
\mciteSetBstMidEndSepPunct{\mcitedefaultmidpunct}
{\mcitedefaultendpunct}{\mcitedefaultseppunct}\relax
\EndOfBibitem
\bibitem[Gordon \latin{et~al.}(2012)Gordon, Fedorov, Pruitt, and
  Slipchenko]{gordon2012fragmentation}
Gordon,~M.~S.; Fedorov,~D.~G.; Pruitt,~S.~R.; Slipchenko,~L.~V. Fragmentation
  methods: A route to accurate calculations on large systems. \emph{Chem. Rev.}
  \textbf{2012}, \emph{112}, 632--672\relax
\mciteBstWouldAddEndPuncttrue
\mciteSetBstMidEndSepPunct{\mcitedefaultmidpunct}
{\mcitedefaultendpunct}{\mcitedefaultseppunct}\relax
\EndOfBibitem
\bibitem[He \latin{et~al.}(2014)He, Zhu, Wang, Liu, and Zhang]{fragzhang}
He,~X.; Zhu,~T.; Wang,~X.; Liu,~J.; Zhang,~J. Z.~H. Fragment Quantum Mechanical
  Calculation of Proteins and Its Applications. \emph{Acc. Chem. Res.}
  \textbf{2014}, \emph{47}, 2748--2757\relax
\mciteBstWouldAddEndPuncttrue
\mciteSetBstMidEndSepPunct{\mcitedefaultmidpunct}
{\mcitedefaultendpunct}{\mcitedefaultseppunct}\relax
\EndOfBibitem
\bibitem[Collins \latin{et~al.}(2014)Collins, Cvitkovic, and
  Bettens]{fragcollins}
Collins,~M.~A.; Cvitkovic,~M.~W.; Bettens,~R. P.~A. The Combined Fragmentation
  and Systematic Molecular Fragmentation Methods. \emph{Acc. Chem. Res.}
  \textbf{2014}, \emph{47}, 2776--2785\relax
\mciteBstWouldAddEndPuncttrue
\mciteSetBstMidEndSepPunct{\mcitedefaultmidpunct}
{\mcitedefaultendpunct}{\mcitedefaultseppunct}\relax
\EndOfBibitem
\bibitem[Qu and Bowman(2019)Qu, and Bowman]{QuBowman2019}
Qu,~C.; Bowman,~J.~M. A Fragmented, Permutationally Invariant Polynomial
  Approach for Potential Energy Surfaces of Large Molecules: Application to
  N-methyl acetamide. \emph{J. Chem. Phys.} \textbf{2019}, \emph{150},
  141101\relax
\mciteBstWouldAddEndPuncttrue
\mciteSetBstMidEndSepPunct{\mcitedefaultmidpunct}
{\mcitedefaultendpunct}{\mcitedefaultseppunct}\relax
\EndOfBibitem
\bibitem[Qu \latin{et~al.}(2024)Qu, Houston, Allison, Schneider, and
  Bowman]{jctctwists}
Qu,~C.; Houston,~P.~L.; Allison,~T.; Schneider,~B.~I.; Bowman,~J.~M. DFT-Based
  Permutationally Invariant Polynomial Potentials Capture the Twists and Turns
  of \ce{C14H30}. \emph{J. Chem. Theory Comput.} \textbf{2024}, \emph{20},
  9339--9353\relax
\mciteBstWouldAddEndPuncttrue
\mciteSetBstMidEndSepPunct{\mcitedefaultmidpunct}
{\mcitedefaultendpunct}{\mcitedefaultseppunct}\relax
\EndOfBibitem
\bibitem[Mor{\'e}(1978)]{more2006levenberg}
Mor{\'e},~J.~J. The Levenberg-Marquardt algorithm: implementation and theory.
  Numerical Analysis. Berlin, Heidelberg, 1978; pp 105--116\relax
\mciteBstWouldAddEndPuncttrue
\mciteSetBstMidEndSepPunct{\mcitedefaultmidpunct}
{\mcitedefaultendpunct}{\mcitedefaultseppunct}\relax
\EndOfBibitem
\bibitem[Wang \latin{et~al.}(2018)Wang, Zhang, Han, \latin{et~al.}
  others]{wang2018deepmd}
Wang,~H.; Zhang,~L.; Han,~J.; others DeePMD-kit: A deep learning package for
  many-body potential energy representation and molecular dynamics.
  \emph{Comput. Phys. Commun.} \textbf{2018}, \emph{228}, 178--184\relax
\mciteBstWouldAddEndPuncttrue
\mciteSetBstMidEndSepPunct{\mcitedefaultmidpunct}
{\mcitedefaultendpunct}{\mcitedefaultseppunct}\relax
\EndOfBibitem
\bibitem[Wiener(1930)]{Wiener1930}
Wiener,~N. Generalized Harmonic Analysis. \emph{Acta Mathematica}
  \textbf{1930}, \emph{55}, 117--258\relax
\mciteBstWouldAddEndPuncttrue
\mciteSetBstMidEndSepPunct{\mcitedefaultmidpunct}
{\mcitedefaultendpunct}{\mcitedefaultseppunct}\relax
\EndOfBibitem
\bibitem[Khintchine(1934)]{Khinchin1934}
Khintchine,~A. {Korrelationstheorie der station\"aren stochastischen Prozesse}.
  \emph{Mathematische Annalen} \textbf{1934}, \emph{109}, 604--615\relax
\mciteBstWouldAddEndPuncttrue
\mciteSetBstMidEndSepPunct{\mcitedefaultmidpunct}
{\mcitedefaultendpunct}{\mcitedefaultseppunct}\relax
\EndOfBibitem
\bibitem[Allen and Tildesley(2017)Allen, and Tildesley]{Allen2017}
Allen,~M.~P.; Tildesley,~D.~J. \emph{Computer simulation of liquids}, 2nd ed.;
  Oxford University Press: Oxford, UK, 2017\relax
\mciteBstWouldAddEndPuncttrue
\mciteSetBstMidEndSepPunct{\mcitedefaultmidpunct}
{\mcitedefaultendpunct}{\mcitedefaultseppunct}\relax
\EndOfBibitem
\bibitem[Qu \latin{et~al.}(2024)Qu, Houston, Conte, and Bowman]{jpcac14h30}
Qu,~C.; Houston,~P.~L.; Conte,~R.; Bowman,~J.~M. Dynamics calculations of the
  flexibility and vibrational spectrum of the linear alkane \ce{C14H30}, based
  on machine-learned potentials. \emph{J. Phys. Chem. A} \textbf{2024},
  \emph{128}, 10713--10722\relax
\mciteBstWouldAddEndPuncttrue
\mciteSetBstMidEndSepPunct{\mcitedefaultmidpunct}
{\mcitedefaultendpunct}{\mcitedefaultseppunct}\relax
\EndOfBibitem
\bibitem[Habershon \latin{et~al.}(2009)Habershon, Markland, and
  Manolopoulos]{q-TIP4P}
Habershon,~S.; Markland,~T.~E.; Manolopoulos,~D.~E. Competing quantum effects
  in the dynamics of a flexible water model. \emph{J. Chem. Phys.}
  \textbf{2009}, \emph{131}, 024501\relax
\mciteBstWouldAddEndPuncttrue
\mciteSetBstMidEndSepPunct{\mcitedefaultmidpunct}
{\mcitedefaultendpunct}{\mcitedefaultseppunct}\relax
\EndOfBibitem
\bibitem[Li \latin{et~al.}(2014)Li, Li, and Ma]{fragli}
Li,~S.; Li,~W.; Ma,~J. Generalized Energy-Based Fragmentation Approach and Its
  Applications to Macromolecules and Molecular Aggregates. \emph{Acc. Chem.
  Res.} \textbf{2014}, \emph{47}, 2712--2720\relax
\mciteBstWouldAddEndPuncttrue
\mciteSetBstMidEndSepPunct{\mcitedefaultmidpunct}
{\mcitedefaultendpunct}{\mcitedefaultseppunct}\relax
\EndOfBibitem
\bibitem[Chen \latin{et~al.}(2019)Chen, Ye, Zuo, Zheng, and Ong]{MEGNet}
Chen,~C.; Ye,~W.; Zuo,~Y.; Zheng,~C.; Ong,~S.~P. Graph networks as a universal
  machine learning framework for molecules and crystals. \emph{Chem. Mater.}
  \textbf{2019}, \emph{31}, 3564--3572\relax
\mciteBstWouldAddEndPuncttrue
\mciteSetBstMidEndSepPunct{\mcitedefaultmidpunct}
{\mcitedefaultendpunct}{\mcitedefaultseppunct}\relax
\EndOfBibitem
\bibitem[Batatia \latin{et~al.}(2025)Batatia, Benner, Chiang, Elena,
  Kov{\'a}cs, Riebesell, Advincula, Asta, Avaylon, Baldwin, and
  others.]{MACE-MP-0}
Batatia,~I.; Benner,~P.; Chiang,~Y.; Elena,~A.~M.; Kov{\'a}cs,~D.~P.;
  Riebesell,~J.; Advincula,~X.~R.; Asta,~M.; Avaylon,~M.; Baldwin,~W.~J.;
  others. A foundation model for atomistic materials chemistry. \emph{J. Chem.
  Phys.} \textbf{2025}, \emph{163}, 184110\relax
\mciteBstWouldAddEndPuncttrue
\mciteSetBstMidEndSepPunct{\mcitedefaultmidpunct}
{\mcitedefaultendpunct}{\mcitedefaultseppunct}\relax
\EndOfBibitem
\end{mcitethebibliography}

\clearpage

\begin{figure}
    \centering
    \includegraphics[width=0.4\linewidth]{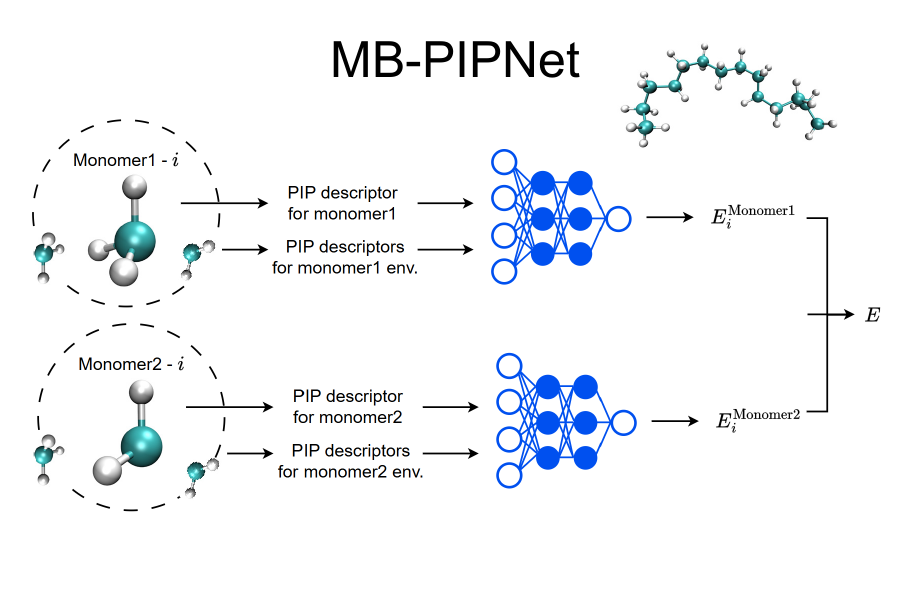}
    \caption*{For Graphical Abstract Only}
    \label{fig:enter-label}
\end{figure}

\end{document}